\documentclass[seceq,preprint]{ptptex}

\usepackage{amsmath}
\usepackage{amssymb}
\usepackage[dvipdfmx]{graphicx}

\preprintnumber{IU-TH-17}

\title{
Formalism of a harmonic oscillator in the future-included complex action theory 
}

\author{%
Keiichi \textsc{Nagao}\footnote{E-mail: keiichi.nagao.phys@vc.ibaraki.ac.jp}
and Holger Bech \textsc{Nielsen}\footnote{E-mail: hbech@nbi.dk}
}

\inst{
$^{*)}$Faculty of Education, Ibaraki University, Bunkyo 2-1-1, Mito 310-8512 
Japan \\
$^{**)}$Niels Bohr Institute, University of Copenhagen, 
Blegdamsvej 17, Copenhagen $\O$, Denmark
}

\abst{%
In a special representation of complex action theory 
that we call ``future-included'', 
we study a harmonic oscillator model defined with 
a non-normal Hamiltonian $\hat{H}$, in which a mass $m$ and an 
angular frequency $\omega$ are taken to be complex numbers. 
In order for the model to be sensible some restrictions on $m$ and $\omega$ are required. 
We draw a phase diagram in the plane of the arguments of $m$ and $\omega$, 
according to which the model is classified into several types. 
In addition, we formulate two pairs of annihilation and creation operators, 
two series of eigenstates of the Hamiltonians 
$\hat{H}$ and $\hat{H}^\dag$, and coherent states. 
They are normalized in a modified inner product $I_Q$, with respect to which  
the Hamiltonian $\hat{H}$ becomes normal. 
Furthermore, applying to the model the maximization principle 
that we previously proposed, 
we obtain an effective theory described by a 
Hamiltonian that is $Q$-Hermitian, i.e. Hermitian with respect to the modified inner product $I_Q$. 
The generic solution to the model is found to be the ``ground'' state. 
Finally we discuss what the solution implies. 
}

\begin{document}

\maketitle


\section{Introduction}

The Feynman path integral (FPI) is a very nice framework for formulating quantum theory. 
We usually consider a real action in the FPI. 
However, if we pursue a fundamental theory, it is better to require fewer conditions 
imposed on it at first. 
Indeed, there is a possibility that the action is complex at the fundamental level but looks real effectively. 
We pursue such  a complex action theory (CAT), which is 
preferable to the usual real action theory (RAT) 
in the sense that the former has at least one fewer conditions: 
there is no reality condition on the action. 
The CAT has been investigated 
with the expectation that the imaginary part of the action 
would give some falsifiable predictions\cite{Bled2006,Nielsen:2008cm,Nielsen:2007ak,Nielsen:2005ub},  
and various interesting suggestions have been made for the Higgs mass\cite{Nielsen:2007mj}, 
quantum-mechanical philosophy\cite{newer1,Vaxjo2009,newer2}, 
some fine-tuning problems\cite{Nielsen2010qq,degenerate}, 
black holes\cite{Nielsen2009hq}, 
de Broglie--Bohm particles, and a cut-off in loop diagrams\cite{Bled2010B}. 
In addition, in Ref.~\cite{Nagao:2010xu}, 
introducing a modified inner product $I_Q$\footnote{Similar inner products are also studied in Refs.~\cite{Geyer, Mostafazadeh_CPT_ip_2002, Mostafazadeh_CPT_ip_2003}.} 
so that a given non-normal Hamiltonian\footnote{The set of non-normal Hamiltonians is much larger 
than that of the PT-symmetric non-Hermitian Hamiltonians, which has been intensively 
studied in Refs.~\cite{Bender:1998ke,Bender:1998gh,Bender:2011ke,Mostafazadeh_CPT_ip_2002,
Mostafazadeh_CPT_ip_2003}. } 
becomes normal with respect to it, 
we proposed a mechanism to effectively 
obtain a Hamiltonian that is $Q$-Hermitian, i.e. 
Hermitian with respect to the modified inner product $I_Q$, after a long time development. 
Furthermore, using the complex coordinate formalism~\cite{Nagao:2011za}, 
we explicitly derived the momentum relation $p=m \dot{q}$, where $m$ is a complex mass, 
via the FPI~\cite{Nagao:2011is}.

The CAT can be classified into two types.  
One is the future-not-included theory~\cite{Nagao:2013eda}, i.e. 
the theory in which the past state $| A(T_A) \rangle$ at the initial time $T_A$ is given, and 
the time integration is performed over the past time.  
The other one is the future-included theory\cite{Bled2006}, 
in which not only the past state 
but also the future state $| B(T_B) \rangle$ at the final time $T_B$ is given at first, 
and the time integration is performed over the whole period from the past to the future. 
In Ref.~\cite{Nagao:2017ecx} we pointed out that 
if a theory is described with a complex action, then such a theory 
is suggested to be the future-included theory 
rather than the future-not-included theory, as long as we respect objectivity. 
In the future-included theory, 
the normalized matrix element~\cite{Bled2006}\footnote{$\langle \hat{\cal O} \rangle^{BA}$ is called the weak value~\cite{AAV} 
in the context of the future-included RAT, and it has been studied intensively. 
The details are found in Ref.~\cite{review_wv} and references therein.}
\begin{equation}
\langle \hat{\cal O} \rangle^{BA} 
\equiv \frac{ \langle B(t) |  \hat{\cal O}  | A(t) \rangle }{ \langle B(t) | A(t) \rangle } , \label{OBA}
\end{equation} 
where $t$ is an arbitrary time ($T_A \leq t \leq T_B$), 
is a strong candidate for the expectation value of an operator $\hat{\cal O}$. 
Indeed, if we regard $\langle \hat{\cal O} \rangle^{BA}$ 
as an expectation value in the future-included theory, 
we obtain the Heisenberg equation, Ehrenfest's theorem, 
and a conserved probability current density~\cite{Nagao:2012mj,Nagao:2012ye}. 
In Ref.~\cite{Nagao:2015bya}, changing the notation of $\langle B(t)|$ 
as $\langle B(t)| \rightarrow \langle B(t)|_Q \equiv \langle B(t)|Q$ 
in $\langle \hat{\cal O} \rangle^{BA}$, 
where 
$Q$ is a Hermitian operator that is appropriately chosen to define the modified inner product $I_Q$, 
we introduced a slightly modified normalized matrix element 
$\langle \hat{\cal O} \rangle_Q^{BA} 
\equiv \frac{ \langle B(t) |_Q  \hat{\cal O}  | A(t) \rangle }{ \langle B(t) |_Q A(t) \rangle }$. 
We proposed a theorem which states that, 
provided that an operator $\hat{\cal O}$ is $Q$-Hermitian, 
$\langle \hat{\cal O} \rangle_Q^{BA}$ becomes real and 
time-develops under a $Q$-Hermitian Hamiltonian for the future and past states 
selected such that the absolute value of the transition amplitude defined with $I_Q$ 
from the past state to the future state is maximized.
We call this way of thinking the maximization principle. 
This theorem 
was proven in both the CAT~\cite{Nagao:2015bya} and the RAT~\cite{Nagao:2017cpl},  
and briefly reviewed in Refs.~\cite{Nagao:2017book, Nagao:2017ztx}.

Through various works explained above we have studied the idea 
that the fundamental action for the universe 
could be complex instead of being real, as is usually assumed. 
A major result of ours is that with regard to the observation 
of the time development there is approximately no deviation 
from what the usual RAT would give, and thus 
there could a priori be the CAT in nature without having immediately seen it. 
The most remarkable deviation from the RAT 
that the CAT predicts is a kind of restriction on the initial conditions. 
Hence we could say that it unifies initial conditions and equations of motion or 
usual quantum mechanics. 
These predictions, however, depend on the detail of the action, 
which has to be guessed as usual. 
To truly settle what type of prediction the CAT leads to, 
a combination of investigation of what the CAT will do 
and guessing of the action to choose is needed. 
To reach the understanding thus required, 
it must be useful to study some examples in the CAT. 
The simplest example from which we can hopefully learn the most important features 
of the CAT is a harmonic oscillator. 
Therefore, in this paper, we shall develop the formalism of the harmonic oscillator 
with parameters $m$ and $\omega$ taken to be complex so that the action becomes complex. 
Even though harmonic oscillators have of course been studied so intensively 
that there is not much chance to do anything new on them, 
we could claim that, since one normally considers it only sensible to work with 
a real action or a Hermitian Hamiltonian, 
we study a seemingly nonsensical and thus not so overstudied theory 
as one a priori thinks about harmonic oscillators. 
Indeed, it would very commonly be assumed that the action is real, and 
in most cases one would neither feel safe 
nor trust studies for the question of the CAT.  
In this sense our work on the harmonic oscillator in the CAT is guaranteed to be new.

Based on the motivation stated above, 
we study the harmonic oscillator model in the future-included CAT. 
After reviewing the complex coordinate formalism~\cite{Nagao:2011za}, 
we provide a non-normal Hamiltonian $\hat{H}$ for the model, 
in which a mass $m$ and an angular frequency $\omega$ are taken to be complex numbers. 
We point out that some restrictions on $m$ and $\omega$ are 
required so that the model becomes sensible. 
According to the argument of $m$ and $\omega$, the model is 
classified into several types. 
We draw a phase diagram in the plane of the arguments of $m$ and $\omega$. 
We formulate two pairs of annihilation 
and creation operators, and 
construct two series of eigenstates $|n \rangle_1$ and 
$|n \rangle_2$ of the Hamiltonians $\hat{H}$ and $\hat{H}^\dag$ respectively 
with several algebraically elegant properties as seen in the usual harmonic oscillator in the RAT. 
Our eigenstates $| n \rangle_1$ and $| n \rangle_2$ are not normalized in a usual sense, but are 
normalized by the condition 
${}_2\langle n | m \rangle_1 = \delta_{nm}$. 
We call this dual normalization. 
In addition, expecting that
classical physics can be described well by coherent states even in the CAT 
as well as in the RAT, 
we construct them for later study.

Next, after reviewing  the modified inner product $I_Q$, with respect to which 
the eigenstates of the Hamiltonian $\hat{H}$ become orthogonal to each other, 
we argue that the dual normalization 
is interpreted as the $Q$-normalization, i.e. the normalization with respect to the inner product $I_Q$. 
Furthermore, we apply the maximization principle to the harmonic oscillator model. 
As a preliminary study, 
supposing that $|A(T_A) \rangle$ and $|B(T_B) \rangle$ are given by the coherent states 
that we constructed, $|\lambda_A (T_A)  \rangle_{\mathrm{coh}, 1}$ and 
$|\lambda_B (T_B)  \rangle_{\mathrm{coh}, 1}$, 
we evaluate $\langle \hat{q}_\mathrm{new} \rangle_Q^{\lambda_B \lambda_A}$ 
and $\langle \hat{p}_\mathrm{new} \rangle_Q^{\lambda_B \lambda_A}$, 
where $\hat{q}_\mathrm{new}$ and $\hat{p}_\mathrm{new}$ are non-Hermitian 
coordinate and momentum operators respectively. 
Then we obtain a classical equation of motion, 
which suggests that, if we obtain a real observable 
$\langle \hat{\cal O} \rangle_Q^{\lambda_B \lambda_A}$ 
via the maximization principle, 
then we have a classical solution, which behaves in a quite similar way to that in the RAT. 
Furthermore, we introduce $Q$-Hermitian coordinate and momentum operators 
$\hat{q}_Q$ and $\hat{p}_Q$, and rewrite the Hamiltonian $\hat{H}$ in terms of 
$\hat{q}_Q$ and $\hat{p}_Q$. 
Utilizing the maximization principle, we obtain an effective theory described by  
a $Q$-Hermitian Hamiltonian that is expressed in terms of $\hat{q}_Q$ and $\hat{p}_Q$.  
We find that the solution to the harmonic oscillator model is the ``ground'' state. 
The ``ground'' state means the state with the utmost energy in the half-infinite series 
of levels. It it only a true ground state for the case of (real) positive $\omega$. 
Finally, we discuss what the solution implies.

This paper is organized as follows. 
In Sect.~\ref{sec:review_complex_coordinate} 
we briefly review the complex coordinate formalism~\cite{Nagao:2011za}. 
In Sect.~\ref{sec:def_ho_phase_diagram} 
we define our harmonic oscillator model and present a phase diagram in the space of 
the arguments of $m$ and $\omega$. 
In Sect.~\ref{sec:two_basis_formalism} we formulate 
two pairs of annihilation and creation operators, and  
construct two series of eigenstates of the Hamiltonians $\hat{H}$ and $\hat{H}^\dag$ 
with the dual normalization. Also, we formulate coherent states. 
In Sect.~\ref{sec:inner_product_IQ}, after reviewing the modified inner product $I_Q$, 
we argue that the dual normalization is interpreted as the normalization with respect to $I_Q$. 
In Sect.~\ref{sec:max_prin_solution_to_ho}, after reviewing the maximization principle, we 
preliminarily study the behavior of $\langle \hat{q}_\mathrm{new} \rangle_Q^{\lambda_B \lambda_A}$ 
and $\langle \hat{p}_\mathrm{new} \rangle_Q^{\lambda_B \lambda_A}$ by supposing that 
$|A(T_A) \rangle$ and $|B(T_B) \rangle$ are given by coherent states 
$|\lambda_A (T_A)  \rangle_{\mathrm{coh}, 1}$ and 
$|\lambda_B (T_B)  \rangle_{\mathrm{coh}, 1}$. 
Finally, we argue that we obtain via the maximization principle an effective theory, 
which is described by a $Q$-Hermitian Hamiltonian, 
and that we are led to the ground state solution. 
Section~\ref{sec:discussion} is devoted to discussion.

\section{Complex coordinate formalism}\label{sec:review_complex_coordinate}


In this section we briefly review the complex coordinate formalism that we proposed 
in Ref.\cite{Nagao:2011za} 
so that we can deal with complex coordinate $q$ and momentum $p$ 
properly not only in the CAT but also in the RAT, 
where we encounter them at the saddle point in the WKB approximation, etc.

\subsection{Non-Hermitian operators 
$\hat{q}_\mathrm{new}$ and $\hat{p}_\mathrm{new}$, 
and the eigenstates of their Hermitian conjugates $|q \rangle_\mathrm{new}$ and 
$|p \rangle_\mathrm{new}$}

We can construct the non-Hermitian operators  of coordinate and momentum, 
$\hat{q}_\mathrm{new}$ and $\hat{p}_\mathrm{new}$, 
and the eigenstates  of their Hermitian conjugates 
$| q \rangle_\mathrm{new}$ and $| p \rangle_\mathrm{new}$, such that 
\begin{eqnarray}
&&\hat{q}_\mathrm{new}^\dag  | q \rangle_\mathrm{new} 
=q | q \rangle_\mathrm{new} , \label{qhatqket=qqket_new} \\
&&\hat{p}_\mathrm{new}^\dag  | p \rangle_\mathrm{new} 
=p | p \rangle_\mathrm{new} , \label{phatpket=ppket_new} \\
&&[\hat{q}_\mathrm{new} , \hat{p}_\mathrm{new}  ] = i \hbar , \label{comqhatphat}
\end{eqnarray}
for complex $q$ and $p$ by formally utilizing two coherent states. 
Our proposal is to replace 
the usual Hermitian operators of coordinate and momentum, 
$\hat{q}$ and $\hat{p}$, and their eigenstates $|q  \rangle$ and $|p  \rangle$, 
which obey 
$\hat{q} | q \rangle = q| q \rangle$, 
$\hat{p} | p \rangle = p | p \rangle$, 
and $[ \hat{q},  \hat{p} ] = i \hbar$ for real $q$ and $p$, 
with $\hat{q}_\mathrm{new}^\dag$, $\hat{p}_\mathrm{new}^\dag$,  
$|q \rangle_\mathrm{new}$, and $|p \rangle_\mathrm{new}$. 
The explicit expressions for $\hat{q}_\mathrm{new}$, $\hat{p}_\mathrm{new}$, 
$| q \rangle_\mathrm{new}$, and $| p \rangle_\mathrm{new}$ are given by 
\begin{eqnarray}
&&\hat{q}_\mathrm{new} \equiv 
\frac{1}{ \sqrt{1 - \epsilon \epsilon' }  } \left( \hat{q} - i \epsilon \hat{p}  \right), \label{def_qhat_new} \\
&&\hat{p}_\mathrm{new} \equiv 
\frac{1}{ \sqrt{1 - \epsilon \epsilon' }  }  \left( \hat{p} + i \epsilon' \hat{q} \right) , \label{def_phat_new} \\
&&| q \rangle_\mathrm{new} 
\equiv 
\left( \frac{1 - \epsilon \epsilon' }{4\pi \hbar \epsilon } \right)^{\frac{1}{4}} 
e^{- \frac{1}{4\hbar\epsilon  } \left( 1 - \epsilon \epsilon' \right) {q}^2 }
| \sqrt{ \frac{1 - \epsilon \epsilon'}{2\hbar \epsilon} } q \rangle_\mathrm{coh} ,  \\
&&| p \rangle_\mathrm{new} 
\equiv 
\left( \frac{1 - \epsilon \epsilon' }{4\pi \hbar \epsilon' } \right)^{\frac{1}{4}} 
e^{ -\frac{1}{4 \hbar \epsilon'}  \left( 1 - \epsilon \epsilon' \right)  p^2 }
| i \sqrt{ \frac{ 1 - \epsilon \epsilon'}{2\hbar \epsilon'} }  p \rangle_\mathrm{coh'} , \label{defpketnew}
\end{eqnarray}
where $| \lambda \rangle_\mathrm{coh}$ is a coherent state parameterized with a complex parameter $\lambda$ defined up to a normalization factor by 
$| \lambda \rangle_\mathrm{coh} 
\equiv  
e^{\lambda {\hat{a}}^\dag} | 0 \rangle 
= \sum_{n=0}^{\infty} \frac{\lambda^n}{\sqrt{n!}} | n \rangle$, 
and this satisfies the relation 
$\hat{a} | \lambda \rangle_\mathrm{coh} = \lambda | \lambda \rangle_\mathrm{coh}$. 
Here, 
$\hat{a} = \sqrt{ \frac{1}{2\hbar \epsilon  }}  \left( \hat{q} + i  \epsilon \hat{p} \right)$ and 
$\hat{a}^\dag = \sqrt{ \frac{1}{2\hbar \epsilon  }}\left( \hat{q} - i  \epsilon \hat{p} \right)$
are annihilation and creation operators. 
In Eq.(\ref{defpketnew}), 
$| \lambda \rangle_\mathrm{coh'} \equiv e^{\lambda {\hat{a'}}^\dag} | 0 \rangle$, 
where ${\hat{a'}}^\dag$  is given by 
${\hat{a'}}^\dag = \sqrt{ \frac{\epsilon'}{2\hbar}}
\left( \hat{q} - i \frac{ \hat{p}}{\epsilon'}  \right) \label{creation'}$, 
is another coherent state defined similarly. 
Before seeing the properties of $\hat{q}_\mathrm{new}$, $\hat{p}_\mathrm{new}$, 
$| q \rangle_\mathrm{new}$, and $| p \rangle_\mathrm{new}$, 
we define a delta function of complex parameters in the next subsection.

\subsection{The delta function}\label{subsec:deltafunc}

\begin{figure}[htb]
\begin{center}
\includegraphics[height=10cm]{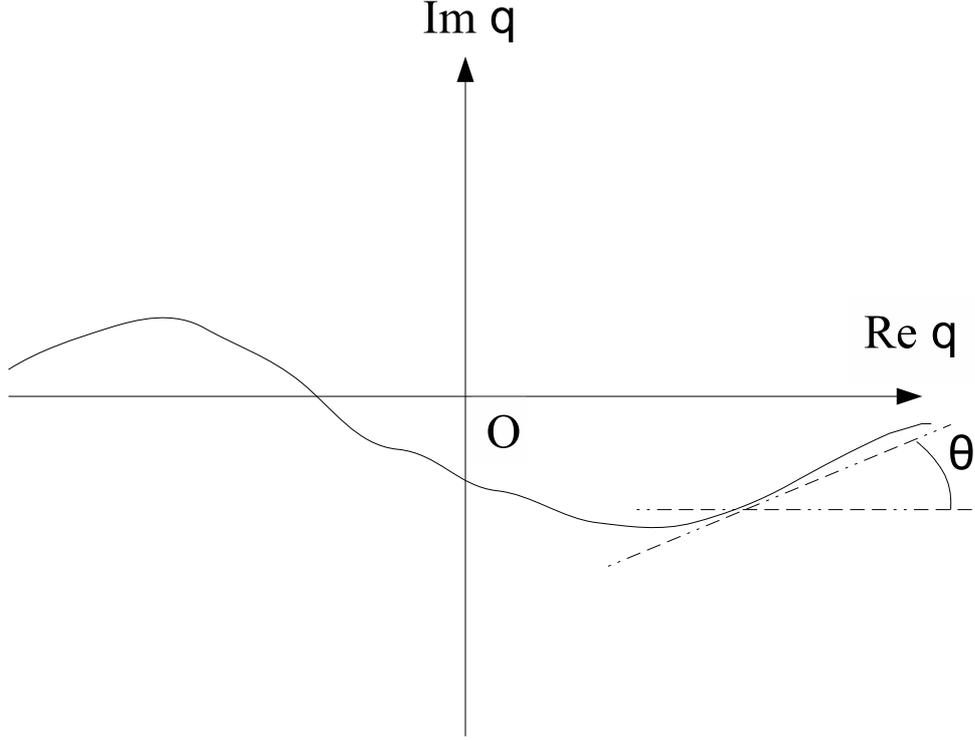}
\end{center}
\caption{Example of a permitted path $C$}
\label{fig:contour}
\end{figure}

\begin{figure}[htb]
\begin{center}
\includegraphics[height=10cm]{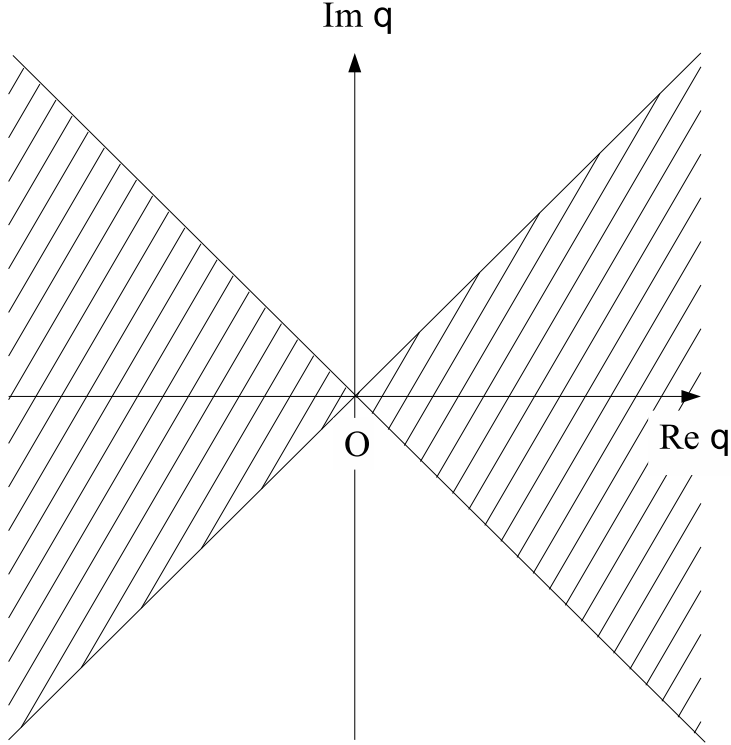}
\end{center}
\caption{Domain of the delta function}
\label{fig:delta_function}
\end{figure}

We define ${\cal D}$ as 
a class of distributions depending on one complex variable $q \in \mathbf{C}$. 
Using a function $g:{\mathbf C} \rightarrow {\mathbf C}$ as 
a distribution\footnote{Another type of complex distribution 
is introduced in Ref.\cite{Nakanishi}. 
It is different from ours in the following points:
the complex distribution in Ref.\cite{Nakanishi}, where $g(q)$ is supposed to have poles, 
is not well defined by $g(q)$ alone, but needs an indication of which side of the 
poles the path $C$ passes through. On the other hand, in our complex distribution 
we assume not the presence of poles of $g(q)$ but $f$ not being a bounded entire 
function.} 
in the class ${\cal D}$, 
we introduce the functional $G[f] = \int_C f(q) g(q) dq$ 
for any analytical function $f:{\mathbf C} \rightarrow {\mathbf C}$ 
with convergence requirements such that $f \rightarrow 0$ for $q \rightarrow \pm \infty$. 
The functional $G$ is a linear mapping from the function $f$ to a complex number. 
Since the simulated function $g$ is supposed to be analytical in $q$, 
the path $C$, which is chosen 
to run from $-\infty$ to $\infty$ in the complex $q$-plane, 
can be deformed freely, and so it is not relevant. 
As an example of such a distribution, we could think of the delta function
and approximate it by 
the smeared delta function defined for complex $q$ by 
\begin{equation}
g(q) = \delta_c^\epsilon(q) 
\equiv \sqrt{\frac{1}{4 \pi \epsilon}} e^{-\frac{q^2}{4\epsilon}} , \label{delta_c_epsilon(q)}
\end{equation}
where $\epsilon$ is a finite small positive real number. 
For the limit of $\epsilon \rightarrow 0$, $g(q)$ behaves as a 
distribution for complex $q$ obeying the condition 
\begin{equation}
L(q) 
\equiv \left( \text{Re}(q) \right)^2 - \left( \text{Im}(q) \right)^2 
>0 . \label{cond_of_q_for_delta} 
\end{equation}
For any analytical test function $f(q)$\footnote{Because of the Liouville theorem, if $f$ is 
a bounded entire function, 
$f$ is constant. So we are considering $f$ as an unbounded entire function or a function 
that is not entire but is holomorphic at least in the region on which the path runs.} 
and any complex $q_0$, this $\delta_c^\epsilon(q)$ satisfies 
$\int_C f(q) \delta_c^\epsilon(q-q_0) dq = f(q_0)$, 
as long as we choose the path $C$ 
such that it runs from $-\infty$ to $\infty$ in the complex $q$-plane 
and at any $q$ its tangent line and a horizontal line 
form an angle $\theta$ whose absolute value  
is within $\frac{\pi}{4}$ to satisfy the inequality in Eq.(\ref{cond_of_q_for_delta}). 
An example of such a permitted path is drawn in Fig.~\ref{fig:contour}. 
Also, the domain of the delta function is shown in Fig.~\ref{fig:delta_function}. 
%


Next, we extend the delta function to complex $\epsilon$, 
and consider 
\begin{eqnarray}
\delta_c^{\epsilon}(aq) 
&=& \sqrt{\frac{1}{4 \pi \epsilon}} e^{-\frac{1}{4\epsilon} a^2 q^2} 
\label{deltaepsilonaq}
\end{eqnarray}
for a non-zero complex $a$. 
We express $\epsilon$, $q$, and $a$ as 
$\epsilon = r_\epsilon e^{i\theta_\epsilon}$, 
$q= r_q e^{i\theta_q}$, and $a = r_a e^{i\theta_a}$. 
The convergence condition of $\delta_c^{\epsilon}(aq)$: 
$\text{Re} \left( \frac{a^2 q^2}{\epsilon} \right) > 0$ 
is expressed as 
\begin{eqnarray}
&&-\frac{\pi}{4} + \frac{1}{2} ( \theta_{\epsilon} - 2\theta_a ) 
< \theta_q < \frac{\pi}{4} + \frac{1}{2} ( \theta_{\epsilon}  - 2 \theta_a ), 
\label{cond1forqa} \\
&&\frac{3}{4} \pi + \frac{1}{2}( \theta_{\epsilon} - 2 \theta_a ) 
< \theta_q < \frac{5}{4}\pi + \frac{1}{2} ( \theta_{\epsilon} - 2 \theta_a ) . 
\label{cond2forqa}
\end{eqnarray}
For $q$, $\epsilon$, and $a$ such that Eqs.(\ref{cond1forqa}) and (\ref{cond2forqa}) 
are satisfied, $\delta_c^{\epsilon}(aq)$ behaves well as a delta function 
of $aq$, and we obtain the relation 
\begin{equation}
\delta_c^{\epsilon}(aq) = 
\frac{\text{sign}(\text{Re}~ a)}{ a } \delta_c^{\frac{\epsilon}{a^2}}(q) , 
\label{scaling_deltafunction} 
\end{equation}
where we have introduced 
\begin{eqnarray}
\text{sign}(\text{Re} a) 
\equiv 
\left\{ 
\begin{array}{cc}
1  & \text{for} ~\text{Re}a > 0 , \\
-1 & \text{for} ~\text{Re}a < 0 . \\ 
\end{array}
\right.
\end{eqnarray}
%

\subsection{New devices to handle complex parameters}\label{newdevices}

To keep the analyticity in dynamical variables of FPI such as 
$q$ and $p$, 
we define a modified set of a complex conjugate, 
real and imaginary parts, bras, and Hermitian conjugates.

\subsubsection{Modified complex conjugate $*_{ \{ \} }$}

We define a modified complex conjugate for a function of $n$ parameters 
$f( \{a_i \}_{i=1, \ldots, n} )$ by 
\begin{equation}
f(\{a_i \}_{i=1, \ldots, n} )^{*_{\{a_i | i \in A \}} } = f^*( \{a_i \}_{i \in A}  ,  \{a_i ^*\}_{i \not\in A} ) , 
\end{equation}
where $A$ denotes the set of indices attached to the parameters 
in which we keep the analyticity, 
and $*$ on $f$ acts on the coefficients included in $f$. 
For example, the complex conjugate 
$*_{q,p}$ 
of a function $f(q,p)=a q^2 + b p^2$ is  
written as $f(q,p)^{*_{q,p}} = a^* q^2 + b^* p^2$.
The analyticity is kept in both $q$ and $p$. 
For simplicity we express the modified complex conjugate as $*_{ \{  \} }$, 
where $\{ \}$ is a symbolic expression for a set of parameters 
in which we keep the analyticity.

\subsubsection{Modified real and imaginary parts $\text{Re}_{\{ \}}$, $\text{Im}_{\{ \}}$ }

We define the modified real and imaginary parts by using $*_{ \{  \} }$.
We decompose some complex function $f$ as 
\begin{equation}
f= \text{Re}_{\{ \}} f + i \text{Im}_{\{ \}} f ,
\end{equation}
where $\text{Re}_{\{ \}} f$ and $\text{Im}_{\{ \}} f$ are 
the ``$\{ \}$-real" and ``$\{ \}$-imaginary" parts of $f$ defined by 
\begin{eqnarray}
&&\text{Re}_{\{ \}} f \equiv \frac{ f + f^{*_{\{ \}}}   }{2} , \label{{}-real} \\
&&\text{Im}_{\{ \}} f \equiv \frac{ f - f^{*_{\{ \}}}   }{2i} . \label{{}-imaginary}
\end{eqnarray}
For example, for $f=k q^2$, the $q$-real and $q$-imaginary parts of $f$ 
are expressed as 
$\text{Re}_{q} f = \text{Re} ( k ) q^2$ and 
$\text{Im}_{q} f = \text{Im} ( k ) q^2$, respectively. 
In particular, if $f$ satisfies $f^{*_{\{ \}}} =f$, 
we say $f$ is $\{ \}$-real, while if $f$ obeys $f^{*_{\{ \}}} =-f$, 
$f$ is purely $\{ \}$-imaginary. 


\subsubsection{Modified bras ${}_m \langle ~|$ and ${}_{ \{ \} } \langle ~|$, 
and modified Hermitian conjugate $\dag_{ \{ \} }$ }

For some state $| \lambda \rangle$ with some complex parameter $\lambda$, 
we define a modified bra ${}_m\langle \lambda |$ by 
\begin{equation} 
{}_m\langle \lambda | \equiv \langle \lambda^* |  \label{modified_bra_anti-linear}
\end{equation}
so that it preserves the analyticity in $\lambda$. 
In the special case of $\lambda$ being real it becomes a normal bra. 
In addition we define a slightly generalized modified bra 
${}_{\{\}}\langle ~|$ and a modified Hermitian conjugate $\dag_{ \{ \} }$ of a ket. 
For example, ${}_{u,v}\langle u | = {}_u \langle u | = {}_m\langle u |$, 
$( | u \rangle )^{\dag_{u, v}}  =( | u \rangle )^{\dag_{u}}  = {}_m \langle u |$. 
We express the Hermitian conjugate $\dag_{ \{ \} }$ of a ket symbolically as 
$( |  ~\rangle )^{\dag_{\{ \} }} = {}_{\{ \}}\langle  ~|$. 
Also, we write the Hermitian conjugate $\dag_{ \{ \} }$ of a bra as 
$( {}_{\{ \}}\langle  ~| )^{\dag_{\{ \} }} =  |  ~\rangle$. 
Hence, for a matrix element we have the relation 
${}_{\{\}}\langle u | A | v \rangle^{*_{ \{ \} }} = 
{}_{ \{ \} } \langle v | A^\dag | u \rangle$.

\subsection{Properties of $\hat{q}_\mathrm{new}$, $\hat{p}_\mathrm{new}$, $|q \rangle_\mathrm{new}$, and $|p \rangle_\mathrm{new}$} \label{prop_qnewpnew}

The states $| q \rangle_\mathrm{new}$ and $| p \rangle_\mathrm{new}$ are normalized 
so that they satisfy 
the following relations: 
\begin{eqnarray}
{}_m\langle_\mathrm{new}~ q' | q \rangle_\mathrm{new} 
&=& \delta_c^{\epsilon_1} ( q' - q ) , \label{m_q'branew_qketnew}\\ 
{}_m\langle_\mathrm{new}~ p' | p \rangle_\mathrm{new} 
&=& \delta_c^{\epsilon'_1} ( p' - p ) , \label{m_p'branew_pketnew}
\end{eqnarray}
where $\epsilon_1$ and $\epsilon'_1$ are given by 
\begin{eqnarray}
\epsilon_1 &\equiv& \frac{\hbar \epsilon }{ 1 -  \epsilon \epsilon' }, \label{epsilon_1} \\
\epsilon'_1 &\equiv& \frac{\hbar \epsilon'}{1 - \epsilon \epsilon' }. \label{epsilon'_1}
\end{eqnarray}
We take $\epsilon$ and $\epsilon'$ sufficiently small, 
for which the delta functions converge 
for complex $q$, $q'$, $p$, and $p'$ satisfying the conditions 
$L(q-q') > 0$ and $L(p-p') > 0$, where $L$ is given in Eq.(\ref{cond_of_q_for_delta}). 
These conditions are satisfied only when $q$ and $q'$ or $p$ and $p'$ are 
on the same paths respectively.  
For small $\epsilon$ and $\epsilon'$, 
Eqs.(\ref{m_q'branew_qketnew}) and (\ref{m_p'branew_pketnew}) represent 
the orthogonality relations for $| q \rangle_\mathrm{new}$ and $| p \rangle_\mathrm{new}$, 
and we have the following relations: 
\begin{eqnarray}
&&\int_C dq | q \rangle_\mathrm{new} ~{}_m \langle_\mathrm{new} q |  \simeq 1 , \label{completion_complexq_ket2} \\
&&\int_C dp | p \rangle_\mathrm{new} ~{}_m \langle_\mathrm{new} p |  \simeq 1 , \label{completion_complexp_ket2} \\
&&\hat{p}_\mathrm{new}^\dag | q \rangle_\mathrm{new} 
\simeq i \hbar \frac{\partial}{\partial q} | q \rangle_\mathrm{new}, \label{phatnewqketnew2} \\
&&\hat{q}_\mathrm{new}^\dag | p \rangle_\mathrm{new} 
\simeq \frac{\hbar}{i} \frac{\partial}{\partial p} | p \rangle_\mathrm{new}, \label{qhatnewpketnew2} \\
&&{}_m\langle_\mathrm{new}~ q | p \rangle_\mathrm{new} 
\simeq \frac{1}{\sqrt{2 \pi \hbar}} \exp\left(\frac{i}{\hbar}p q \right). 
\label{basis_fourier_transf2}  
\end{eqnarray}
Thus, $\hat{q}_\mathrm{new}^\dag$, $\hat{p}_\mathrm{new}^\dag$, 
$| q \rangle_\mathrm{new}$, and $| p \rangle_\mathrm{new}$ 
with complex $q$ and $p$ 
obey the same relations as $\hat{q}$, $\hat{p}$, $| q \rangle$, 
and $| p \rangle$ with real $q$ and $p$. 
In the $\epsilon \rightarrow 0$ and $\epsilon' \rightarrow 0$ limits, 
$\delta_c^{\epsilon_1} ( q' - q )$, $\delta_c^{\epsilon'_1} ( p' - p )$, and 
$\exp\left( \frac{i}{\hbar}p q \right)$ 
in Eqs.(\ref{m_q'branew_qketnew}), (\ref{m_p'branew_pketnew}), and (\ref{basis_fourier_transf2}) 
are well defined as distributions of the class ${\cal D}$. 
For real $q'$ and $p'$, $| q' \rangle_\mathrm{new}$ and $| p' \rangle_\mathrm{new}$ become 
$| q' \rangle$ and $| p' \rangle$ respectively; 
also, $\hat{q}_\mathrm{new}^\dag$ and $\hat{p}_\mathrm{new}^\dag$ behave like  
$\hat{q}$ and $\hat{p}$ respectively.

\section{Harmonic oscillator model and phase diagram in $m$ and $\omega$} \label{sec:def_ho_phase_diagram}

In this section,  
after reviewing the future-included theory, 
we define our harmonic oscillator model in the CAT and present the phase diagram.

\subsection{Harmonic oscillator Hamiltonian in the future-included theory}

\subsubsection{Future-included theory} \label{sec:review_future-included_theory}

The future-included theory\cite{Bled2006,Nagao:2012mj,Nagao:2012ye} is 
described by using 
the future state $| B (T_B) \rangle$ at the final time $T_B$ 
and the past state $| A (T_A) \rangle$ at the initial time $T_A$. 
For a given non-normal Hamiltonian $\hat{H}$, 
$| A (t) \rangle$ and $| B (t) \rangle$ 
obey the Schr\"{o}dinger equations 
\begin{eqnarray}
&&i \hbar \frac{d}{dt} | A(t) \rangle = \hat{H} | A(t) \rangle , \label{schro_eq_Astate} \\
&&
i \hbar \frac{d}{dt} | B(t) \rangle = {\hat{H}}^{\dag} | B(t) \rangle , 
\label{schro_eq_Bstate_old} 
\end{eqnarray}
and are expressed as 
\begin{eqnarray}
| A(t) \rangle
&=& e^{-\frac{i}{\hbar} \hat{H}(t-T_A)} | A(T_A) \rangle  , \label{Atket} \\
| B(t) \rangle  
&=& e^{-\frac{i}{\hbar} \hat{H}^\dag (t-T_B)} | B(T_B) \rangle . \label{Btket}
\end{eqnarray}
In Refs.\cite{Nagao:2012mj,Nagao:2012ye}, we investigated the normalized matrix element 
$\langle \hat{\cal O} \rangle^{BA} 
\equiv \frac{ \langle B(t) |  \hat{\cal O}  | A(t) \rangle }{ \langle B(t) | A(t) \rangle }$, 
which is called the weak value\cite{AAV, review_wv} in the RAT, 
and found that if we regard $\langle \hat{\cal O} \rangle^{BA}$ 
as an expectation value in the future-included theory, 
then we obtain the Heisenberg equation, Ehrenfest's theorem, 
and a conserved probability current density. 
In fact, since $\langle \hat{\cal O} \rangle^{BA}$ obeys 
\begin{eqnarray}
\frac{d}{dt} \langle \hat{\cal O} \rangle^{BA} 
&=& \langle  \frac{i}{\hbar} [ \hat{H} , \hat{\cal O} ]   \rangle^{BA}  
\label{time_dev_OBA}
\end{eqnarray} 
for a general Hamiltonian 
\begin{equation}
\hat{H}=\frac{1}{2m} \hat{p}_\mathrm{new}^2 + V(\hat{q}_\mathrm{new}), \label{generalH}
\end{equation}
where $V$ is a general potential defined by $V(q)=\sum_{n=2}^\infty b_n q^n$, we obtain 
\begin{eqnarray}%
&&\frac{d}{dt} \langle \hat{q}_\mathrm{new} \rangle^{BA} 
= \frac{1}{m} \langle \hat{p}_\mathrm{new} \rangle^{BA} ,  \label{dqcdt}  \\
&&\frac{d}{dt} \langle \hat{p}_\mathrm{new} \rangle^{BA} 
= - \langle V'(\hat{q}_\mathrm{new}) \rangle^{BA} ,    \label{dpcdt} 
\end{eqnarray}
and Ehrenfest's theorem, 
$m\frac{d^2}{dt^2} \langle \hat{q}_\mathrm{new} \rangle^{BA} 
= - \langle V'(\hat{q}_\mathrm{new}) \rangle^{BA}$. 
Thus, $\langle \hat{\cal O} \rangle^{BA}$ provides 
the time development of the saddle point for $\exp(\frac{i}{\hbar} S)$,  
and seems to have the role of an expectation value in the future-included theory. 
In addition, let us introduce a probability density $\rho$ by 
\begin{equation}
\rho 
\equiv \frac{\psi_B(q)^{*_q} \psi_A(q)}{\langle B | A \rangle} 
=  \frac{ \langle B | q \rangle_\mathrm{new}  ~{}_m\langle_\mathrm{new}~ q | A \rangle }{ \langle B | A \rangle } , 
\end{equation}
which satisfies $\int_C dq \rho =1$, 
where $C$ is an arbitrary contour running from $-\infty$ to $\infty$ 
in the complex $q$-plane. 
Then we can construct a conserved probability current density $j$ by 
\begin{equation}
j(q,t) \equiv  \frac{ \frac{i\hbar}{2m} \left(  \frac{\partial \tilde{\psi}_B^{*_q}  }{\partial q}  
\psi_A - \tilde{\psi}_B^{*_q}   \frac{\partial \psi_A }{\partial q} \right) }{ \langle B | A \rangle } , 
\end{equation}
which obeys the continuity equation 
$\frac{\partial \rho}{\partial t } +  \frac{\partial }{\partial q} j(q,t) = 0$. 
Therefore, probability interpretation seems to work formally with this $\rho$.

As for the Lagrangian, in Ref.~\cite{Nagao:2011is}, 
starting from the Hamiltonian given in Eq.(\ref{generalH}), 
we obtained via the FPI the Lagrangian 
$L(q, \dot{q})=\frac{1}{2}m \dot{q}^2- \sum_{n=2}^\infty b_n q^n$, and vice versa. 
In addition, we derived via the FPI the momentum relation 
\begin{equation}
p(t)=m \frac{d}{d t} q(t) . \label{p=mqdot}
\end{equation}
We note that this is not the case in the future-not-included CAT. 
Indeed, we showed in Ref.~\cite{Nagao:2013eda} that in the future-not-included CAT 
the Lagrangian and momentum relation are given by 
$L_{\text{eff}}(\dot{q}, q) 
= \frac{1}{2} m_{\text{eff}}~ \dot{q}^2 - \sum_{n=2}^\infty \text{Re} b_n~ q^n$ 
and $p=m_\mathrm{eff} \dot{q}$, 
where $m_{\text{eff}} \equiv m_\mathrm{R} + \frac{m_\mathrm{I}^2}{m_\mathrm{R}}$. 
Since Eq.(\ref{dqcdt}) is consistent with Eq.(\ref{p=mqdot}), Eq.(\ref{p=mqdot}) 
is confirmed to be the momentum relation in the future-included theory.

\subsubsection{Harmonic oscillator Hamiltonian}

Utilizing $\hat{q}_\mathrm{new}$ and $\hat{p}_\mathrm{new}$ given in Eqs.(\ref{def_qhat_new}) and (\ref{def_phat_new}), 
we define our harmonic oscillator Hamiltonian $\hat{H}$ by 
\begin{eqnarray}
&&\hat{H} 
\equiv \frac{1}{2m} \hat{p}_\mathrm{new}^2 + V(\hat{q}_\mathrm{new}) , \label{hoHamiltonian} \\
&&V(\hat{q}_\mathrm{new}) = \frac{1}{2}m\omega^2 \hat{q}_\mathrm{new}^2 , \label{ho_potential_hat} 
\end{eqnarray}
where both mass $m$ and angular frequency $\omega$ are complex, and decomposed as follows: 
\begin{eqnarray}
&&m=m_\mathrm{R} + i m_\mathrm{I} = r_m e^{i \theta_m} ,  \label{m} \\
&&\omega= \omega_\mathrm{R} + i \omega_\mathrm{I} = r_\omega e^{i \theta_\omega} , \label{omega}
\end{eqnarray}
where $m_\mathrm{R}$, $\omega_\mathrm{R}$, $m_\mathrm{I}$, and $\omega_\mathrm{I}$ are the real and imaginary parts of 
$m$ and $\omega$, and  
$r_m$, $r_\omega$, $\theta_m$, and $\theta_\omega$ are the absolute values and 
arguments of $m$ and $\omega$, respectively. 
This Hamiltonian depends on $\epsilon$ and $\epsilon'$ 
via $\hat{q}_\mathrm{new}$ and $\hat{p}_\mathrm{new}$. 
For our later convenience, let us introduce another Hamiltonian that is 
independent of $\epsilon$ and $\epsilon'$, 
\begin{equation}
\hat{H}_{\epsilon=\epsilon'=0} \equiv \frac{1}{2m} \hat{p}^2 + \frac{1}{2} m \omega^2 \hat{q}^2 , 
\label{H_epsilon=0}
\end{equation}
by taking the limits $\epsilon \rightarrow 0$ and $\epsilon' \rightarrow 0$, or 
replacing $\hat{q}_\mathrm{new}$ and $\hat{p}_\mathrm{new}$ with $\hat{q}$ and $\hat{p}$ in $\hat{H}$. 
Utilizing the fact 
obtained in Ref.~\cite{Nagao:2011is},  
we find that the Lagrangian is simply given by 
\begin{eqnarray}
L(q, \dot{q}) &=& \frac{1}{2}m \dot{q}^2- V(q) , \label{lagrangian} \\
V(q) &=& \frac{1}{2}m\omega^2 q^2 . \label{ho_potential}
\end{eqnarray}
The potential $V$ is decomposed as 
\begin{eqnarray} 
V&=&V_\mathrm{R} + iV_\mathrm{I} , \\
V_\mathrm{R}
&\equiv&\text{Re}_q V 
=\text{Re} \left( \frac{m\omega^2}{2} \right) q^2 , \label{V_R}  \\
V_\mathrm{I}
&\equiv&\text{Im}_q V 
=\text{Im} \left( \frac{m\omega^2}{2} \right) q^2 , \label{V_I} 
\end{eqnarray}
where $\text{Re}_q$ and $\text{Im}_q$ are introduced in Eqs.(\ref{{}-real}) and (\ref{{}-imaginary}).

We consider the functional integral 
$\int_C {\cal D} q~ \psi_B^*  \psi_A e^{\frac{i}{\hbar} \int L(q, \dot{q}) dt }$, 
and suppose that the asymptotic values of dynamical variables such as $q$ and $p$ are 
on the real axis. 
The path $C$ denotes an arbitrary path running from $-\infty$ to $\infty$ in the complex plane 
for each moment of time $t$, 
and we can deform it as long as the integrand keeps the analyticity in $q$ and $p$. 
To prevent the kinetic term in the integrand 
from blowing up for $\dot{q}\rightarrow \pm\infty$ along the real axis, 
we impose on $m$ the condition\footnote{In an exact sense, the convergent condition 
is given by $m_\mathrm{I} > 0$, while we know that the harmonic oscillator model 
with $m_\mathrm{I}=0$ works well in the RAT. 
Hence we have included $m_\mathrm{I}=0$ for the condition in Eq.(\ref{mIgeq0}). 
Similarly, we have included $\text{Im} (m \omega^2) =0$ for the condition in Eq.(\ref{Immomega2leq0}). 
Note that if $m_\mathrm{I}$ or $\text{Im} (m \omega^2)$ violated 
the two conditions in Eqs.(\ref{mIgeq0}) and (\ref{Immomega2leq0}), i.e. 
if  $m_\mathrm{I} < 0$ or $\text{Im} (m \omega^2) > 0$, 
then the functional integral divergence would be exponential, and thus 
it would be much more serious than the divergence trouble in the RAT, where 
$m_\mathrm{I} = 0$ and $\text{Im} (m \omega^2) = 0$. } 
\begin{equation}
m_\mathrm{I} \geq 0 .   \label{mIgeq0}
\end{equation} 
In addition, to ensure the convergence of the functional integral, 
we need the following condition on the potential: 
\begin{equation}
\text{Im} (m \omega^2) \leq 0 .  \label{Immomega2leq0}
\end{equation}
Then, since 
$m\omega$ and $m\omega^2$ are written as 
\begin{eqnarray} 
&&m\omega \equiv r e^{i\theta}=r_m r_\omega  e^{i ( \theta_m + \theta_\omega ) }, \label{momega} \\
&&m\omega^2 =  r_m r_\omega^2 e^{i ( \theta_m + 2\theta_\omega)} ,  \label{momega2} 
\end{eqnarray}
the two conditions in Eqs.(\ref{mIgeq0}) and (\ref{Immomega2leq0}) 
are expressed in terms of $\theta_m$ and $\theta_\omega$ as 
\begin{eqnarray}
&& 0 \leq \theta_m \leq \pi , \label{mIgeq02}   \\
&& - \pi \leq \theta_m + 2\theta_\omega \leq 0 
\quad \leftrightarrow \quad 
-\frac{\theta_m}{2} -\frac{\pi}{2} \leq \theta_\omega \leq - \frac{\theta_m}{2} ,  
\label{Immomega2leq02}  
\end{eqnarray}
respectively. 
%

\subsection{Study of the phase diagram}

In this subsection we analyze the phase diagram in the $(\theta_m, \theta_\omega)$ plane. 
We will see that, according to the values of $\theta_m$ and $\theta_\omega$, 
our harmonic oscillator model includes several different theories. 
Indeed, the value of $\theta_m$ classifies the model into 
the usual time theory (UTT), imaginary time theory (ITT) and flipped time theory (FTT). 
Also, according to the value of $\theta_\omega$, 
not only a harmonic oscillator (HO) but also an inverted harmonic oscillator (IHO) 
is described.

Using Eq.(\ref{momega2}), let us express $V_\mathrm{R}$ and $V_\mathrm{I}$ given in Eqs.(\ref{V_R}) and (\ref{V_I}) as 
\begin{eqnarray} 
V_\mathrm{R}
&=&\frac{q^2}{2} r_m r_\omega^2 
\cos(\theta_m + 2\theta_\omega) , \label{V_Rmomega22}\\
V_\mathrm{I}
&=&\frac{q^2}{2} r_m r_\omega^2 
\sin(\theta_m + 2\theta_\omega) . \label{V_Imomega22}
\end{eqnarray}
Then, according to the signs of $V_\mathrm{R}$ and $V_\mathrm{I}$, 
the permitted region of $\theta_\omega$ by the condition in Eq.(\ref{Immomega2leq02}) 
can be classified into the following five regions: 
\begin{enumerate}
\item For $\theta_\omega =  -\frac{\theta_m}{2} 
\quad\Leftrightarrow\quad  \theta_m + 2\theta_\omega = 0$: 

$V_\mathrm{R} > 0$, $V_\mathrm{I}=0$.  

\item For $-\frac{\theta_m}{2} - \frac{\pi}{4} < \theta_\omega <
  -\frac{\theta_m}{2} \quad\Leftrightarrow\quad   - \frac{\pi}{2} < \theta_m + 2\theta_\omega < 0$: 

$V_\mathrm{R} > 0$, $V_\mathrm{I}<0$. 

\item For $\theta_\omega=-\frac{\theta_m}{2} - \frac{\pi}{4} \quad\Leftrightarrow\quad    \theta_m + 2\theta_\omega =-\frac{\pi}{2}$: 

$V_\mathrm{R} = 0$, $V_\mathrm{I}<0$. 

\item For $-\frac{\theta_m}{2} - \frac{\pi}{2} <  \theta_\omega < -\frac{\theta_m}{2} - \frac{\pi}{4} \quad\Leftrightarrow\quad   -\pi < \theta_m + 2\theta_\omega < -\frac{\pi}{2}$: 

$V_\mathrm{R} < 0$, $V_\mathrm{I} <0$.  

\item For $\theta_\omega=-\frac{\theta_m}{2} - \frac{\pi}{2}  \quad\Leftrightarrow\quad  \theta_m + 2\theta_\omega =-\pi$: 

$V_\mathrm{R} < 0$, $V_\mathrm{I}=0$. 
\end{enumerate} 
Later, using the different condition in Eq.(\ref{mIgeq02}), 
we investigate these regions in more detail according to the value of $\theta_m$.

\subsubsection{Our principle of interpretation of various quantities in the CAT}\label{subsubsec:principle_interpretation}

We shall explain our interpretation of various quantities in the CAT. 
We allow both mass $m$ and angular frequency $\omega$ to be complex, 
so negative numbers are naturally included. Since we have a much larger class of theories, 
there can only be a priori less chance that we obtain just what we find in nature.  
Some possible outcomes will simply disagree with some of our experiences. 
We have to choose the parameters appropriately. 
We then divide the possibilities for the sign of the real part of $m$ called $m_\mathrm{R}$ 
to classify the theories. 
We think that the real part of (non-relativistic) mass should be positive 
in a sensible theory. 
One possible strategy would be to declare that there is an empirical law that $m_\mathrm{R}$ shall be positive. 
Another one would be to introduce some transformation 
to change the mass into a new mass so that its real part becomes positive. 
Based on this way of thinking\footnote{ It might be also reasonable to think that 
the real part of the angular frequency $\omega$ should be positive. 
If we take this philosophy for $\omega$, or take both the philosophies for $m$ and $\omega$,  
then the harmonic oscillator model could be classified in slightly different ways. 
However, in this paper we elucidate the phase structure of the harmonic oscillator model 
only by taking the philosophy for $m$ for simplicity. }, 
we define a new mass by 
\begin{equation}
m_\mathrm{new} \equiv a m , \label{mnewam}
\end{equation}
where $a$, whose magnitude is $1$, is properly chosen so that $\text{Re}~m_\mathrm{new} > 0$. 
Since $\theta_m = \arg m$ is restricted by the condition in Eq.(\ref{mIgeq02}), 
$a$ is chosen according to the sign of $m_\mathrm{R}$, as shown later.

Next we introduce new times $t_\mathrm{new}$ and $T_A^\mathrm{new}$, and 
a new angular frequency $\omega_\mathrm{new}$ 
by demanding the relation 
\begin{eqnarray} 
\exp\left[-\frac{i}{\hbar} 
\hat{H} (t-T_A) \right] 
&=& 
\exp\left[ -\frac{i}{\hbar} 
\hat{H}_\mathrm{new}
 (t_\mathrm{new}-T_A^\mathrm{new})  \right]  \label{exp[-ioverhbarHhat(t-TA)_new]}
\end{eqnarray}
for the Hamiltonian 
$\hat{H}$ given in Eqs.(\ref{hoHamiltonian}) and (\ref{ho_potential_hat}), 
and a new Hamiltonian $\hat{H}_\mathrm{new}$ defined by 
\begin{eqnarray}
\hat{H}_\mathrm{new} 
&\equiv& 
\frac{\hat{p}_\mathrm{new}^2}{2m_\mathrm{new}} +  \frac{1}{2}m_\mathrm{new} \omega_\mathrm{new}^2 \hat{q}_\mathrm{new}^2 
=\frac{1}{a} \hat{H} .  \label{Hnew1overaH} 
\end{eqnarray}
Comparing the free parts of $\hat{H}$ and $\hat{H}_\mathrm{new}$ 
on both sides of Eq.(\ref{exp[-ioverhbarHhat(t-TA)_new]}), we define 
\begin{equation}
t_\mathrm{new} \equiv \frac{m_\mathrm{new}}{m} t = at , \label{t_new} 
\end{equation}
and $T_A^\mathrm{new} \equiv \frac{m_\mathrm{new}}{m} T_A = a T_A$. 
Similarly, we define $T_B^\mathrm{new} \equiv a T_B$. 
In addition, we introduce a new pair of coordinate and momentum, 
$q_\mathrm{new}$ and $p_\mathrm{new}$, by 
\begin{eqnarray}
q_\mathrm{new}(t_\mathrm{new}) &\equiv&q(t) , \label{qnew=q} \\
p_\mathrm{new}(t_\mathrm{new}) &\equiv& p(t). \label{pnew=p}
\end{eqnarray}
Using  Eqs.(\ref{t_new})-(\ref{pnew=p}), 
we can rewrite the momentum relation given in Eq.(\ref{p=mqdot})  
in terms of the new variables as 
$p_\mathrm{new}(t_\mathrm{new}) 
= m_\mathrm{new} \frac{d}{d t_\mathrm{new}} q_\mathrm{new}(t_\mathrm{new})$. 
Next we compare the potential terms of $\hat{H}$ and $\hat{H}_\mathrm{new}$ 
on both sides of Eq.(\ref{exp[-ioverhbarHhat(t-TA)_new]}). 
Then we might feel like defining 
$\omega_\mathrm{new} = \pm \frac{1}{a} \omega$, 
where we encounter an indefiniteness for the sign of $\omega_\mathrm{new}$.  
However, since the expression of Eq.(\ref{Hnew1overaH}) suggests a new energy 
$E_\mathrm{new} \equiv \frac{1}{a} \lambda_n$, 
if we suppose that we can obtain an energy eigenvalue 
$\lambda_n \equiv \hbar \omega \left( n + \frac{1}{2} \right)$\footnote{We obtain 
the same energy eigenvalue in Eq.(\ref{lambda_n}) of 
Sect.~\ref{sec:annihilation_creation_operators}. } for $\hat{H}$, 
we are led to defining $\omega_\mathrm{new}$ with a definite sign by 
\begin{equation}
\omega_\mathrm{new} \equiv \frac{1}{a} \omega ,  \label{omeganew}
\end{equation}
so that $E_\mathrm{new}$ is expressed as 
$E_\mathrm{new} =\hbar \omega_\mathrm{new} \left( n + \frac{1}{2} \right)$. 
Equation (\ref{omeganew}) is also given by demanding the relation 
$\omega t = \omega_\mathrm{new} t_\mathrm{new}$.

According to the sign of $m_\mathrm{R}$, we determine 
$m_\mathrm{new}$, $\omega_\mathrm{new}$, and $t_\mathrm{new}$ as follows: 
\begin{enumerate}
\item For $0 \leq \theta_m < \frac{\pi}{2}$: 

Since $m_\mathrm{R}>0$, we choose $a=1$, i.e. $m_\mathrm{new}= m$, $\omega_\mathrm{new}=\omega$, and $t_\mathrm{new}=t$. 

\item For $\theta_m = \frac{\pi}{2}$: 

Since $m_\mathrm{R} =0$, we choose $a=-i$, i.e.  $m_\mathrm{new}=-im$, $\omega_\mathrm{new}=i\omega$, 
and $t_\mathrm{new}=-it$.

\item For $\frac{\pi}{2} < \theta_m \leq \pi$: 

Since $m_\mathrm{R}<0$, we choose $a=-1$, i.e.  $m_\mathrm{new}= -m$, $\omega_\mathrm{new}=-\omega$, and $t_\mathrm{new}=-t$. 
\end{enumerate}
Unless one transforms the negativity of $m_\mathrm{R}$ away, 
cases 2 and 3 would be forbidden by the empirical law that $m_\mathrm{R}$ shall be positive.

\subsubsection{The phase diagram}\label{subsubsec:phase_diagram}

\begin{figure}[htb]
\begin{center}
\includegraphics[height=9cm]{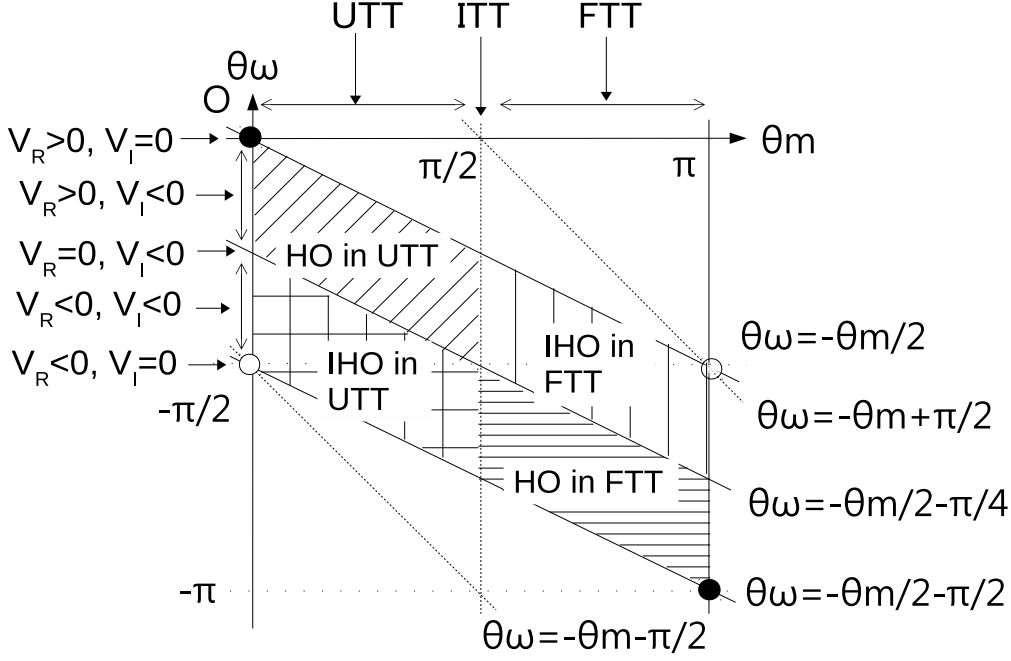}
\end{center}
\caption{The phase diagram of the harmonic oscillator defined with complex $m$ and $\omega$. 
Here $\theta_m=\arg m$ and $\theta_\omega=\arg \omega$, as defined in 
Eqs.(\ref{m}) and (\ref{omega}). The special cases contained in the RAT 
are at the four corners of the allowed parallelogram region. A usual harmonic oscillator model having 
positive energy is defined at the upper left corner. }
\label{fig:phase_diagram_ho}
\end{figure}

Based on the strategy given in Sect.~\ref{subsubsec:principle_interpretation}, 
we can classify 
our harmonic oscillator model into several theories. 
We have presented such an explicit study in Appendix~\ref{appendix:classification}. 
Thus, the phase diagram of the harmonic oscillator specified by 
Eqs.(\ref{mIgeq02}) and (\ref{Immomega2leq02}) is drawn 
in Fig.\ref{fig:phase_diagram_ho}\footnote{For our later convenience 
to consider the condition in Eq.(\ref{momegatheta_cond}) for there being 
eigenstates of $\hat{H}$ and coherent states in Sect.~\ref{sec:two_basis_formalism}, 
the two lines $\theta_\omega = -\theta_m \pm \frac{\pi}{2}$ have also been drawn. 
The investigation in the following sections, based mainly 
on the two-basis formalism of eigenvectors forming ladder states, 
is valid in the whole parallelogram region allowed by 
Eqs.(\ref{mIgeq02}) and (\ref{Immomega2leq02}) 
except for the two corners $(\theta_m, \theta_\omega)=(0, - \frac{\pi}{2}), (\pi, - \frac{\pi}{2})$, 
which are not allowed by the condition in Eq.(\ref{momegatheta_cond}). 
The two corners represent inverse harmonic oscillators in the RAT. }.  
%

\section{Two-basis formalism} \label{sec:two_basis_formalism}

In this section we develop our two-basis formalism of eigenvectors 
for the harmonic oscillator Hamiltonians 
$\hat{H}$ and $\hat{H}^\dag$.

\subsection{Annihilation and creation operators}\label{sec:annihilation_creation_operators}

We define two annihilation operators, $\hat{a}_1$ and $\hat{a}_2$, and creation operators, 
$\hat{a}_1^\dag$ and $\hat{a}_2^\dag$, by their Hermitian conjugates as follows: 
\begin{eqnarray}
&&\hat{a}_1 = \sqrt{\frac{m\omega}{2 \hbar}} \left(\hat{q}_\mathrm{new} +\frac{i\hat{p}_\mathrm{new} }{m\omega} \right) ,
\label{a_1} \\
&&\hat{a}_2 = 
\sqrt{\frac{m^* \omega^*}{2 \hbar}} \left(\hat{q}_\mathrm{new} ^\dag 
+\frac{i\hat{p}_\mathrm{new} ^\dag }{m^* \omega^*} \right) , \label{a_2}
\\
&&\hat{a}_2^\dag  = \sqrt{\frac{m\omega}{2 \hbar}} \left(\hat{q}_\mathrm{new}  -\frac{i\hat{p}_\mathrm{new}  }{m\omega} \right) , \label{a_2_dag} \\
&&\hat{a}_1^\dag = 
\sqrt{\frac{m^* \omega^*}{2 \hbar}} 
\left(\hat{q}_\mathrm{new} ^\dag -\frac{i\hat{p}_\mathrm{new} ^\dag }{m^* \omega^*} \right). 
\label{a_1_dag} 
\end{eqnarray}
Equations (\ref{a_1}) and (\ref{a_2_dag}) provide $\hat{q}_\mathrm{new} $ and $\hat{p}_\mathrm{new}$ 
in terms of $\hat{a}_1$ and $\hat{a}_2^\dag$ as 
\begin{eqnarray}
\hat{q}_\mathrm{new}  
&=& \sqrt{ \frac{\hbar}{ 2 m \omega} } (\hat{a}_1 + \hat{a}_2^\dag) , \label{newqhat} \\
\hat{p}_\mathrm{new}  
&=& -i \sqrt{ \frac{\hbar m \omega}{2}} (\hat{a}_1 - \hat{a}_2^\dag). \label{newphat}
\end{eqnarray}
Then, the commutation relation $[ \hat{q}_\mathrm{new} , \hat{p}_\mathrm{new} ] = i\hbar$ is written 
as 
\begin{equation}
[ \hat{a}_1, \hat{a}_2^\dag] = 1, \label{comm_a_1_a_2dag}
\end{equation}
and the Hamiltonian $\hat{H}$ in Eq.(\ref{hoHamiltonian}) and its Hermitian conjugate 
$\hat{H}^\dag$ are expressed in terms of $\hat{a}_1$ and $\hat{a}_2^\dag$ as 
\begin{eqnarray} 
\hat{H} 
&=& \hbar \omega \left( \hat{a}_2^\dag \hat{a}_1 + \frac{1}{2} \right) , \\
\hat{H}^\dag 
&=&  \hbar \omega^* \left( \hat{a}_1^\dag \hat{a}_2 + \frac{1}{2} \right) . 
\end{eqnarray}
We define two ground states $| 0 \rangle_1$ and $| 0 \rangle_2$ up to the normalization by 
\begin{eqnarray}
&&\hat{a}_1 | 0 \rangle_1 = 0 , \label{a10ket1=0} \\ 
&&\hat{a}_2 | 0 \rangle_2 = 0 , \label{a20ket2=0}
\end{eqnarray}
and excited states $| n \rangle_1$ and $| n \rangle_2$ for positive integer $n$ 
up to the normalization as 
$| n \rangle_1 \propto (\hat{a}_2^\dag )^n | 0 \rangle_1$ and 
$| n \rangle_2 \propto (\hat{a}_1^\dag )^n | 0 \rangle_2$. 
In addition, we introduce number operators $\hat{n}_1$ and $\hat{n}_2$ by 
\begin{eqnarray}
&&\hat{n}_1=\hat{a}_2^\dag \hat{a}_1 , \label{nhat1} \\
&&\hat{n}_2=\hat{a}_1^\dag \hat{a}_2 = \hat{n}_1^\dag . \label{nhat2} 
\end{eqnarray}
Then they obey 
$\hat{n}_1| n \rangle_{1} = n  | n \rangle_{1}$ and $\hat{n}_2| n \rangle_{2} = n  | n \rangle_{2}$, 
and $\hat{H}$ and $\hat{H}^\dag$ are expressed as 
\begin{eqnarray}
&&\hat{H} = \hbar \omega \left( \hat{n}_1 + \frac{1}{2} \right) , \label{H_hbaromega_n} \\
&&\hat{H}^\dag = \hbar \omega^* \left( \hat{n}_2 + \frac{1}{2} \right) . \label{Hdagger_hbaromega*_n}
\end{eqnarray}
We see that 
$| n \rangle_{1}$ and $| n \rangle_{2}$ are eigenstates of $\hat{H}$ and $\hat{H}^\dag$, 
\begin{eqnarray}
&&\hat{H} | n \rangle_{1} = \hbar \omega \left( n + \frac{1}{2} \right) | n \rangle_{1} , 
\label{Hhatnket1} \\
&&\hat{H}^\dag | n \rangle_{2} = \hbar \omega^* \left( n + \frac{1}{2} \right) | n \rangle_{2} , 
\end{eqnarray}
so, in particular, $\hat{H}$ has the following eigenvalue for $| n \rangle_{1}$:
\begin{equation}
\lambda_n = \hbar \omega \left( n+ \frac{1}{2} \right) . \label{lambda_n} 
\end{equation}
Here we note that $| n \rangle_{1}$ and $| n \rangle_{2}$ are not orthogonal 
eigenstates; ${}_1\langle m | n \rangle_1$ and ${}_2\langle m | n \rangle_2$ 
are not proportional to $\delta_{mn}$, 
since $\hat{H}$ and $\hat{H}^\dag$ are not Hermitian. 
Though these eigenstates $| n \rangle_1$ and $| n \rangle_{2}$ are technically somewhat hard 
to normalize, we can construct rather easily two series of eigenstates 
that are not genuinely normalized 
but fixed by a convention that makes the algebra of $\hat{a}_2^\dag$ and $\hat{a}_1$ 
work very elegantly like in the RAT case.

\subsection{Normalization of $| n \rangle_1$ and $| n \rangle_{2}$} \label{subsec:norm_nket1_nket2}

In this subsection we shall discuss 
how we normalize the series of Hilbert vectors $| n \rangle_1$ and $| n \rangle_{2}$. 
There could be a number of ways of normalizing them. 
We first explain them.

1) We can imagine the special set of $| n \rangle_1$ 
by a naive analytical continuation of  the $q$-representation of 
the normalized state in the RAT, 
$| n \rangle = \frac{1}{\sqrt{n !}} (\hat{a}^\dag )^n | 0 \rangle$, to complex $m\omega$ for small 
$\epsilon$ and $\epsilon'$: 
\begin{equation}
{}_m\langle_\mathrm{new}~ q | n \rangle_1 
\simeq 
\left( \frac{m\omega}{\pi \hbar} \right) ^\frac{1}{4}
\frac{1}{\sqrt{n!}} 
\left(   \frac{1}{ \sqrt{2}}  \right)^n
H_n \left( \sqrt{\frac{m \omega}{\hbar}} q  \right)
\exp\left(  -\frac{m\omega}{2 \hbar}  q^2 \right)  , \label{HPS}
\end{equation}
where on the left-hand side 
we have used a modified bra for complex $q$, 
and on the right-hand side $H_n(x)$ is the $n$th Hermite polynomial, 
$H_n(x) = \exp\left( \frac{1}{2} x^2 \right)  \left( x - \frac{d}{dx}  \right)^n 
\exp\left( -\frac{1}{2} x^2 \right)$. 
In particular, ${}_m\langle_\mathrm{new}~ q | 0 \rangle_1$ is expressed as 
\begin{equation}
{}_m\langle_\mathrm{new}~ q | 0 \rangle_1 
\simeq \left(  \frac{m\omega}{\pi \hbar} \right)^{ \frac{1}{4} } 
\exp\left( - \frac{m\omega}{2\hbar} q^2 \right).  \label{qbra0ket1} 
\end{equation}
Replacing $m\omega$ with $m^* \omega^*$ in the RAT state $| n \rangle$ 
and then analytically continuing in $m^* \omega^*$, we obtain 
the set $| n \rangle_{2}$ for small $\epsilon$ and $\epsilon'$: 
\begin{eqnarray}
{}_m\langle_\mathrm{new}~ q |n \rangle_{2} 
&\simeq&
\left( \frac{m^* \omega^*}{\pi \hbar} \right) ^\frac{1}{4}
\frac{1}{\sqrt{n!}} 
\left(   \frac{1}{ \sqrt{2}}  \right)^n
H_n \left( \sqrt{\frac{m^* \omega^*}{\hbar}} q  \right)
\exp\left(  -\frac{m^* \omega^*}{2 \hbar}  q^2 \right).    
\label{HPS2}
\end{eqnarray}
Let us consider the correction to complex $q$ for the $n$th Hermite polynomial $H_n(q)$. 
$H_n (q)$ is a smooth $q$-wave function for small $n$, but not so for large $n$, 
for which it oscillates considerably. 
Comparing the expressions for the Hamiltonian $\hat{H}$ 
in Eqs.(\ref{hoHamiltonian}) and (\ref{Hhatnket1}), we see that 
$q$ and $p$ classically go up in proportion to $\sqrt{n}$ for large $n$. 
Hence, the width of $H_n (q)$ is proportional to $\sqrt{n}$. 
In addition, $H_n (q)$ has $n$ zeros. 
Since the density of zeros is about $\frac{n}{\sqrt{n}}=\sqrt{n}$ per unit length in $q$,  
the length of each wave contained in $H_n (q)$ is about $\frac{1}{\sqrt{n}}$. 
On the other hand, the correction to complex $q$ is $\epsilon p \sim \epsilon \sqrt{n}$. 
It is $\epsilon \sqrt{n}/ {\frac{1}{\sqrt{n}}} \sim \epsilon n$ relative to the wave length. 
Therefore, when $\epsilon n >1$ we cannot ignore the $\epsilon p$ term anymore. 
So the expressions in Eqs.(\ref{HPS}) and (\ref{HPS2}) are valid for $n$ 
such that $n < \frac{1}{\epsilon}$.

The expression of Eq.(\ref{HPS}), 
which is a function of $m \omega$ but not $m^* \omega^*$, 
motivates us to define our $| n \rangle_1$ including the factor in front by 
\begin{equation}
| n \rangle_1 \equiv \frac{1}{\sqrt{n !}} (\hat{a}_2^\dag )^n | 0 \rangle_1.  \label{nket1normalized}
\end{equation}
The state $| n \rangle_1$ is not normalized in the usual sense. 
The squared norm of $| n \rangle_1$ involves both $m\omega$ and $m^* \omega^*$, 
so it is not analytic in $m\omega$. 
Similarly, we are motivated to define our $| n \rangle_2$ by 
\begin{equation}
| n \rangle_2 \equiv \frac{1}{\sqrt{n !}} (\hat{a}_1^\dag )^n | 0 \rangle_2. \label{nket2normalized}
\end{equation}

2) We could also single out our proposed series of eigenstate $| n \rangle_1$ 
by the requirement of the usual ladder formulas 
with $\hat{a}_2^\dag$ and $\hat{a}_1$ replacing $a^\dag$ and $a$ respectively, 
\begin{eqnarray}
&&\hat{a}_2^\dag | n \rangle_1 = \sqrt{n+1}  | n +1  \rangle_1 , \label{a2dagstep}\\
&&\hat{a}_1 | n \rangle_1 = \sqrt{n}  | n - 1  \rangle_1 . \label{a1step}
\end{eqnarray}
This algebraic requirement -- not involving any norm -- specifies the $| n \rangle_1$ state 
even with respect to $n$-dependent scale factors. 
To consider the set $| n \rangle_2$ in the same way, 
the algebraic requirement in Eqs.(\ref{a2dagstep}) and (\ref{a1step}) 
should be replaced with the following ladder equations: 
\begin{eqnarray}
&&\hat{a}_1^\dag | n \rangle_2 = \sqrt{n+1}  | n +1  \rangle_2 , \label{a1dagstep} \\
&&\hat{a}_2 | n \rangle_2 = \sqrt{n}  | n - 1  \rangle_2 .  \label{a2step}
\end{eqnarray}
In our definitions $\hat{a}_1^\dag$ and $\hat{a}_2$ are the ladder operators 
depending on $m^* \omega^*$, while 
$\hat{a}_2^\dag$ and $\hat{a}_1$ used for construction of the $| n \rangle_1$ states 
are the ones depending on $m\omega$.

3) The third possibility is to try to determine both the prefactors of $| n \rangle_1$ 
and $| n \rangle_2$ by imposing the condition 
\begin{equation}
{}_2\langle m | n \rangle_1 = \delta_{mn}   \label{2branmket1}
\end{equation}
on $| n \rangle_1$ and $| m \rangle_2$. 
This condition means that $| m \rangle_2$ is regarded as 
a dual basis of $| n \rangle_1$, and also implies the 
following completeness relation: 
\begin{equation}
\sum_{n=0}^\infty | n \rangle_1 ~{}_2\langle n | =1 . \label{sumnnket1n2bra}
\end{equation} 
If we write $| n \rangle_1$ and $| m \rangle_2$ as 
$| n \rangle_1 = C_1(n) (\hat{a}_2^\dag )^n | 0 \rangle_1$ and 
$| m \rangle_2 = C_2(m) (\hat{a}_1^\dag )^m | 0 \rangle_2$, 
then Eq.(\ref{2branmket1}) gives only the condition 
$C_2(n)^* C_1(n) = \frac{1}{n!}$. 
Choosing $C_1(n)$ and $C_2(n)$ symmetrically as 
$C_1(n) = C_2(n)=\frac{1}{ \sqrt{ n!} }$ 
leads to the $| n \rangle_1$ of Eq.(\ref{nket1normalized}) specified by $1)$ and $2)$, 
and the analogue for $| n \rangle_2$ given in Eq.(\ref{nket2normalized}). 
This procedure $3)$ does not quite fix the normalization of $| n \rangle_1$ 
alone, but needs to be supplemented by $1)$ or $2)$. 
The condition in Eq.(\ref{2branmket1}) 
indeed follows from the scale specifications suggested under $1)$ and $2)$, i.e. 
the analytical continuation and the ladder relation requirements respectively,  
if they are supplemented by the analogous construction of the $| n \rangle_2$ states. 
We call this ``dual normalization".

Using the above rules $1)$, $2)$, and $3)$, which are consistent with each other, we have specified 
two series of eigenstates $| n \rangle_1$ and $| n \rangle_2$ of $\hat{H}$ and $\hat{H}^\dag$ 
respectively. They formally look like being normalized in the usual sense, but actually 
only in the sense of the dual normalization by Eq.(\ref{2branmket1}). 
The two-basis formalism of $| n \rangle_1$ and $| m \rangle_2$ is our replacement for 
the usual formalism of $| n \rangle$ in the RAT. 
Indeed, we first define our ground states $| 0 \rangle_1$ and $| 0 \rangle_2$ by 
Eqs.(\ref{a10ket1=0}), (\ref{a20ket2=0}), and (\ref{2branmket1}), where we choose 
their normalization factors symmetrically\footnote{In Appendix~\ref{appendix:ground_states}, 
we give concrete expressions for $| 0 \rangle_1$ and $| 0 \rangle_2$.}. 
Second, we define our $| n \rangle_1$ and $| n \rangle_2$ for $n \ge 1$ by 
Eqs.(\ref{nket1normalized}) and (\ref{nket2normalized}). 
Then we obtain for the overlap ${}_2\langle m | n \rangle_1$ 
the same result $\delta_{mn}$ as in the RAT, i.e. Eq.(\ref{2branmket1}), 
and our states $| n \rangle_1$ and $| n \rangle_2$ obey the ladder relations given in 
Eqs.(\ref{a2dagstep}), (\ref{a1step}), (\ref{a1dagstep}), and (\ref{a2step}).

The point is that, when we take the bra ${}_2\langle m |$ 
correlated to the ket $| m \rangle_2$, 
we get an expression formally written in terms of $m\omega$, 
and thus the overlap ${}_2\langle m | n \rangle_1$ 
becomes an integral of an expression involving only $m \omega$ 
to be an analytical continuation of 
$\langle m | n \rangle$ in $m\omega$, which is well known to give $\delta_{mn}$. 
For ${}_{2} \langle n | n' \rangle_1$ 
we can see this property even 
by using the concrete expressions of Eqs.(\ref{HPS}) and (\ref{HPS2}) 
for small $\epsilon$ and $\epsilon'$ as follows:  
\begin{eqnarray} 
{}_{2} \langle n | n' \rangle_1
&\simeq&
\int dq ~{}_{2} \langle n | q \rangle_\mathrm{new} ~{}_m \langle_\mathrm{new}~ q |n' \rangle_1 \nonumber \\
&\simeq&
\frac{1}{\sqrt{n ! n' ! }} \left( \frac{1}{\sqrt{2}} \right)^{n+n'} 
\left( \frac{m \omega}{\pi \hbar} \right) ^\frac{1}{2}
\left( \frac{\hbar}{m \omega} \right) ^\frac{1}{2}
\int dX H_n(X) H_{n'}(X) \exp(-X^2) \nonumber \\
&=& 
\delta_{n n'} , \label{n2bran'ket1}
\end{eqnarray}
where in the second line we have changed the variable $q$ into 
$X=\sqrt{\frac{m \omega}{\hbar}} q=\sqrt{\frac{r}{\hbar}} e^{i \frac{\theta}{2}} q$, 
where $r$ and $\theta$ are introduced in Eq.(\ref{momega}). 
In the last equality, we have used 
the following relation for complex $X$ 
by rotating the integration contour by the angle $|\frac{\theta}{2} |$: 
$\int_{-\infty}^{\infty} dX H_n(X) H_{n'}(X) e^{-X^2} = \sqrt{\pi} 2^n n!  \delta_{n n'}$, 
which is valid for $\theta$ such that 
\begin{eqnarray}
&&|\theta| < \frac{\pi}{2}   
~\leftrightarrow~ 
-\theta_m - \frac{\pi}{2} < \theta_\omega < -\theta_m + \frac{\pi}{2} . \label{momegatheta_cond}
\end{eqnarray}
Therefore, this is the condition for $| n \rangle_1$ and $| n \rangle_2$ to be normalizable 
in the sense of Eq.(\ref{2branmket1}). 
If, however, we ask for overlaps of $| n \rangle_1$ states 
with each other, ${}_1\langle m | n \rangle_1$, 
or those of $| n \rangle_2$ states with each other, ${}_2\langle m | n \rangle_2$, 
then, since $| n \rangle_1$ and $|n \rangle_2$ 
are not normalized in the usual inner product, 
we obtain overlap integrals 
with both $m\omega$ and $m^* \omega^*$ 
appearing formally. 
These integrals are not simple analytical continuations of the RAT integrals. 
In Sect.~\ref{choice_Q} we will show that 
the dual normalization by Eq.(\ref{2branmket1}) 
can be regarded as an orthonormal condition of $|n\rangle_1$ 
or $|n\rangle_2$ with respect to an inner product $I_Q$ or $I_{Q^{-1}}$ 
defined there, respectively.

\subsection{Coherent states made of $| n \rangle_1$ and $| n \rangle_{2}$} 
\label{subsec:coherent_states_made_of_nket1_nket2}

It is strongly suggested that if we want to see classical dynamics 
of a harmonic oscillator, we should study coherent states. 
Indeed, in the RAT coherent states are thought to be 
classical states represented by wave packets, so we now attempt to 
construct coherent states in the CAT. 
We utilize one of the coherent states 
in Sect.~\ref{subsec:ABstates_being_coherent_states}.

Following the two-basis formalism developed in the previous subsections, 
we define two coherent states $| \lambda \rangle_{\mathrm{coh}, 1}$ and 
$| \lambda \rangle_{\mathrm{coh}, 2}$ by 
\begin{eqnarray}
&& 
| \lambda \rangle_{\mathrm{coh}, 1} 
= e^{-\frac{|\lambda|^2}{2}} e^{\lambda \hat{a}_2^\dag} | 0 \rangle_1 
=\sum_{n=0}^{\infty} f(n) |n \rangle_1 , \label{lambdaket1} \\
&& 
| \lambda \rangle_{\mathrm{coh}, 2}
= e^{-\frac{|\lambda|^2}{2}} e^{\lambda \hat{a}_1^\dag} | 0 \rangle_2 
=\sum_{n=0}^{\infty} f(n) |n \rangle_2 , \label{lambdaket2} 
\end{eqnarray}
where $f(n)$ is given by 
\begin{equation}
f(n) = e^{-\frac{|\lambda|^2}{2}} \frac{\lambda^n}{\sqrt{n!}} . \label{fn}
\end{equation}
Here, the coefficients $e^{-\frac{|\lambda|^2}{2}}$ of the center expressions of 
Eqs.(\ref{lambdaket1}) and (\ref{lambdaket2}) are chosen symmetrically 
so that in the RAT limit 
$| \lambda \rangle_{\mathrm{coh}, 1}$ and $| \lambda \rangle_{\mathrm{coh}, 2}$ 
have the same forms as the coherent state in the RAT. 
The two coherent states satisfy 
\begin{eqnarray}
&&\hat{a}_1 | \lambda \rangle_{\mathrm{coh}, 1} =  \lambda | \lambda \rangle_{\mathrm{coh}, 1} , 
\label{a1lambdaketcoh1} \\
&&\hat{a}_2 | \lambda \rangle_{\mathrm{coh}, 2} =  \lambda | \lambda \rangle_{\mathrm{coh}, 2} , 
\label{a2lambdaketcoh2} 
\end{eqnarray}
which can be checked by using the relations 
$[\hat{a}_1, (\hat{a}_2^\dag)^n ] =  n (\hat{a}_2^\dag)^{n-1}$, 
$[\hat{a}_1, e^{\lambda \hat{a}_2^\dag} ] =  \lambda e^{\lambda \hat{a}_2^\dag}$, 
$[\hat{a}_2, (\hat{a}_1^\dag)^n ] =  n (\hat{a}_1^\dag)^{n-1}$, and 
$[\hat{a}_2, e^{\lambda \hat{a}_1^\dag} ] =  \lambda e^{\lambda \hat{a}_1^\dag}$.  
Since the overlap of $| \lambda_B \rangle_{\mathrm{coh},2}$ 
and $| \lambda_A \rangle_{\mathrm{coh},1}$ is given by 
${}_{\mathrm{coh}, 2} \langle \lambda_B | \lambda_A \rangle_{\mathrm{coh}, 1} 
=\exp\left[
-\frac{1}{2}\left(  |\lambda_B|^2 - 2 \lambda_B^* \lambda_A + |\lambda_A|^2 \right)
\right]$, 
they are normalized by  
${}_{\mathrm{coh}, 2}\langle \lambda | \lambda \rangle_{\mathrm{coh}, 1} 
= {}_2\langle 0 | 0 \rangle_1 = 1$, 
and obey 
$\frac{1}{\pi} \int d^2 \lambda 
| \lambda \rangle_{\mathrm{coh}, 1} ~{}_{\mathrm{coh}, 2}\langle \lambda | = 
\sum_{n=0}^{\infty} | n \rangle_1 ~{}_2\langle n |=1$, 
where $d^2 \lambda  = d\lambda_R d\lambda_I$.

Incidentally, we give the $q$-representation of the coherent state 
$| \lambda \rangle_{\mathrm{coh}, 1}$ for small $\epsilon$ and $\epsilon'$.  
For this purpose we utilize the relation 
\begin{equation}
e^{\lambda \hat{a}_2^\dag} =
\exp\left(  \lambda \sqrt{\frac{m\omega}{2 \hbar} }  \hat{q}_\mathrm{new}  \right)
\exp\left(  -i \lambda \sqrt{\frac{ 1}{ 2 \hbar m \omega}  }  \hat{p}_\mathrm{new}   \right)
e^{-\frac{1}{4} \lambda^2} , \label{elambdaadag}
\end{equation}
which can be derived by using Eq.(\ref{a_2_dag}) and 
$e^{\hat{A} + \hat{B}} = e^{\hat{A}} e^{\hat{B}}  e^{ -\frac{1}{2} [\hat{A}, \hat{B} ] }$, 
which holds for operators $\hat{A}$ and $\hat{B}$ such that 
$[\hat{A}, \hat{B} ]$ is a classical number. 
Then the $q$-representation of the coherent state $| \lambda \rangle_{\mathrm{coh}, 1}$ 
for small $\epsilon$ and $\epsilon'$ is given by 
\begin{eqnarray}
{}_m\langle_\mathrm{new}~ q | \lambda  \rangle_{\mathrm{coh}, 1} 
&\simeq&e^{-\frac{|\lambda|^2}{2}} 
e^{-\frac{1}{4} \lambda^2}
\exp\left(  \lambda \sqrt{\frac{m\omega}{2 \hbar} }  q \right) 
~{}_m\langle_\mathrm{new}~ q -\lambda \sqrt{\frac{ \hbar}{ 2 m \omega}  }  | 0 \rangle_1 \nonumber \\
&\simeq&
e^{\frac{1}{2} ( \lambda^2 - |\lambda|^2 ) } 
\left( \frac{m\omega}{\pi \hbar} \right) ^\frac{1}{4}
\exp\left[  -\frac{m\omega}{2 \hbar}  
\left( q  -\lambda \sqrt{\frac{ 2 \hbar}{ m \omega}  } \right)^2 \right] , 
\label{qbralamdaketcoh1}
\end{eqnarray} 
where in the first equality we have used 
Eqs.(\ref{lambdaket1}) and (\ref{elambdaadag}), and in the second equality 
we have used Eq.(\ref{qbra0ket1}). 
Equation (\ref{qbralamdaketcoh1}) suggests that 
for the coherent state $| \lambda  \rangle_{\mathrm{coh}, 1}$ to be normalizable we need 
the following condition on $m \omega$: 
\begin{equation}
\text{Re}(m \omega) > 0 . 
\end{equation}
This is the same as the condition in Eq.(\ref{momegatheta_cond}) 
for $| n \rangle_1$ and $| n \rangle_2$ 
to be normalizable in the sense of Eq.(\ref{2branmket1}). 
Similarly, we obtain the $q$-representation of the coherent state 
$| \lambda \rangle_{\mathrm{coh}, 2}$ for small $\epsilon$ and $\epsilon'$: 
\begin{eqnarray}
~{}_m\langle_\mathrm{new}~ q | \lambda  \rangle_{\mathrm{coh}, 2} 
&\simeq&
e^{\frac{1}{2} ( \lambda^2 - |\lambda|^2 ) } 
\left( \frac{m^* \omega^*}{\pi \hbar} \right) ^\frac{1}{4}
\exp\left[  -\frac{m^* \omega^*}{2 \hbar}  
\left( q  -\lambda \sqrt{\frac{ 2 \hbar}{ m^* \omega^*} } 
\right)^2 \right] . 
\label{qbralamdaketcoh2}
\end{eqnarray} 
The condition for the coherent state $| \lambda \rangle_{\mathrm{coh}, 2}$ to be normalizable 
is the same as in Eq.(\ref{momegatheta_cond}).

In the phase diagram shown in Fig.\ref{fig:phase_diagram_ho} 
we have seen that some phases have a healthy real part, but 
others even violate the positivity of the Hermitian part of the Hamiltonian. 
Nevertheless, our treatment with the two-basis formalism 
will be applicable as long as the ground states are achievable. 
We note that 
the condition in Eq.(\ref{momegatheta_cond}) excludes the two corners 
$(\theta_m, \theta_\omega)=(\pi, - \frac{\pi}{2}), (0, - \frac{\pi}{2})$ 
from the parallelogram region permitted by Eqs.(\ref{mIgeq02}) and (\ref{Immomega2leq02}).  
Therefore, our treatment extends to the whole parallelogram 
except for the two corners in the phase diagram. 
The two troublesome corners represent inverted harmonic oscillators in the RAT. 
Indeed, their kinetic terms $T$ and potential terms $V$ go oppositely: 
one has $T \ge 0$ and $V \le 0$, while the other has $T \le 0$ and $V \ge 0$.


We summarize various quantities of our two-basis formalism 
in Table~\ref{tab1:two-basis}.
\begin{table}
\caption[AAA]{Summary of the two-basis formalism for the two Hamiltonians 
$\hat{H}$ and $\hat{H}^\dag$}
\label{tab1:two-basis}
\begin{center}
\begin{tabular}{|p{4cm}|p{5cm}|p{6cm}|}
\hline
& For $\hat{H}=\frac{\hat{p}_\mathrm{new}^2}{2m} + \frac{1}{2} m \omega^2 \hat{q}_\mathrm{new}^2$: 
& For $\hat{H}^\dag=\frac{(\hat{p}_\mathrm{new}^\dag)^2}{2m^*} + \frac{1}{2} m^* (\omega^*)^2 (\hat{q}_\mathrm{new}^\dag)^2$: \\
\hline
Annihilation operator   
& 
$\hat{a}_1 = \sqrt{\frac{m\omega}{2 \hbar}} \left(\hat{q}_\mathrm{new} +\frac{i\hat{p}_\mathrm{new} }{m\omega} \right)$ 
& 
$\hat{a}_2 = \sqrt{\frac{m^* \omega^*}{2 \hbar}} \left(\hat{q}_\mathrm{new} ^\dag 
+\frac{i\hat{p}_\mathrm{new} ^\dag }{m^* \omega^*} \right) $ 
\\  
\hline
Creation operator  
& 
$\hat{a}_2^\dag  = \sqrt{\frac{m\omega}{2 \hbar}} \left(\hat{q}_\mathrm{new}  -\frac{i\hat{p}_\mathrm{new}  }{m\omega} \right)$ 
& 
$\hat{a}_1^\dag = 
\sqrt{\frac{m^* \omega^*}{2 \hbar}} 
\left(\hat{q}_\mathrm{new} ^\dag -\frac{i\hat{p}_\mathrm{new} ^\dag }{m^* \omega^*} \right)$ \\  
\hline
Ground state   
& $| 0 \rangle_1$ defined by $\hat{a}_1 | 0 \rangle_1 = 0$ 
& $| 0 \rangle_2$ defined by $\hat{a}_2 | 0 \rangle_2 = 0$ \\  
\hline
$n$-state   
& $| n \rangle_1 = \frac{1}{\sqrt{n !}} (\hat{a}_2^\dag )^n | 0 \rangle_1$ 
& $| n \rangle_2 = \frac{1}{ \sqrt{n!} } (\hat{a}_1^\dag )^n | 0 \rangle_2$ \\  
\hline 
Ladder equation     
& $\hat{a}_1 | n \rangle_1 = \sqrt{n}  | n - 1  \rangle_1$ ,  
& $\hat{a}_2 | n \rangle_2 = \sqrt{n}  | n - 1  \rangle_2$ ,  \\ 
& $\hat{a}_2^\dag | n \rangle_1 = \sqrt{n+1}  | n +1  \rangle_1$ 
& $\hat{a}_1^\dag | n \rangle_2 = \sqrt{n+1}  | n +1  \rangle_2$ \\ 
\hline
Number operator   
& $\hat{n}_1=\hat{a}_2^\dag \hat{a}_1$ , 
& $\hat{n}_2=\hat{a}_1^\dag \hat{a}_2 = \hat{n}_1^\dag$ ,  \\  
& $\hat{n}_1| n \rangle_{1} = n  | n \rangle_{1}$ 
& $\hat{n}_2| n \rangle_{2} = n  | n \rangle_{2}$ \\  
\hline
Commutation relation   
& $[\hat{a}_1 , \hat{a}_2^\dag]=1$ , 
& $[\hat{a}_2 , \hat{a}_1^\dag]=1$ ,  \\  
& $[\hat{n}_1 , \hat{a}_1]= - \hat{a}_1$ , $[\hat{n}_1 , \hat{a}_2^\dag]= \hat{a}_2^\dag$
& $[\hat{n}_2 , \hat{a}_2]= - \hat{a}_2$ , $[\hat{n}_2 , \hat{a}_1^\dag]= \hat{a}_1^\dag$ \\   
\hline
Hamiltonian     
& $\hat{H} = \hbar \omega \left( \hat{n}_1 + \frac{1}{2} \right)$ , 
& $\hat{H}^\dag = \hbar \omega^* \left( \hat{n}_2 + \frac{1}{2} \right)$ , \\  
& $\hat{H} | n \rangle_{1} = \hbar \omega \left( n + \frac{1}{2} \right) | n \rangle_{1}$ 
& $\hat{H}^\dag | n \rangle_{2} = \hbar \omega^* \left( n + \frac{1}{2} \right) 
| n \rangle_{2}$ \\ 
\hline
$q$-representation       
& ${}_m\langle_\mathrm{new}~ q | n \rangle_1 \simeq$ 
& ${}_m\langle_\mathrm{new}~ q | n \rangle_{2} \simeq$ \\ 
of the eigenstate
& $\left( \frac{m\omega}{\pi \hbar} \right) ^\frac{1}{4}
\frac{1}{\sqrt{n!}} 
\left(   \frac{1}{ \sqrt{2}}  \right)^n$ 
& $\left( \frac{m^* \omega^*}{\pi \hbar} \right) ^\frac{1}{4}
\frac{1}{\sqrt{n!}} 
\left(   \frac{1}{ \sqrt{2}}  \right)^n$ \\ 
& $\times H_n \left( \sqrt{\frac{m \omega}{\hbar}} q  \right)
\exp\left(  -\frac{m\omega}{2 \hbar}  q^2 \right)$ 
& $\times H_n \left( \sqrt{\frac{m^* \omega^*}{\hbar}} q  \right)
\exp\left(  -\frac{m^* \omega^*}{2 \hbar}  q^2 \right)$ \\
\hline
\end{tabular}
\end{center}
\end{table}
%
%

\section{On the inner product $I_Q$} \label{sec:inner_product_IQ}

In the previous section we constructed two sets of eigenstates $|n \rangle_1$ and 
$|n \rangle_2$ for the Hamiltonians $\hat{H}$ and $\hat{H}^\dag$ respectively with several algebraically 
elegant properties as seen in the usual harmonic oscillator in the RAT. 
These states $| n \rangle_1$ and $| n \rangle_2$ are not orthogonal to each other. 
They are dual-normalized by Eq.(\ref{2branmket1}), not normalized in the usual sense. 
In this section, after reviewing the modified inner product $I_Q$, 
we argue that the dual normalization of Eq.(\ref{2branmket1}) 
can be interpreted as the normalization condition with respect to the 
inner product $I_Q$.

\subsection{Review of the modified inner product $I_Q$}

It is easy to see that Eq.(\ref{2branmket1}) can be interpreted as a formal orthogonality relation 
provided we introduce the modified inner product $I_Q$  
for arbitrary states $|\psi_1 \rangle$ and $| \psi_2 \rangle$
in the Hilbert space by 
\begin{equation}
I_Q(| \psi_1\rangle, | \psi_2 \rangle) 
\equiv \langle \psi_1 |_Q \psi_2 \rangle 
\equiv \langle \psi_1 | Q | \psi_2 \rangle,  \label{Qin}
\end{equation}
where $Q$ is chosen so that the eigenstates of a given non-normal Hamiltonian $\hat{H}$,  
$| \lambda_i \rangle_1$, which obey 
$\hat{H} | \lambda_i \rangle_1 = \lambda_i | \lambda_i \rangle_1$, 
become orthogonal to each other, 
\begin{equation}
I_Q(| \lambda_i \rangle_1, | \lambda_j \rangle_1) 
 = {}_1\langle \lambda_i |_Q \lambda_j \rangle_1 
= \delta_{ij}. \label{Qin2}
\end{equation}
In Refs.\cite{Nagao:2010xu,Nagao:2011za} 
we put forward the idea of introducing such a modified inner product $I_Q$. 
Then, $\hat{H}$, being not even normal, 
$[\hat{H}^\dag , \hat{H} ] \neq 0$, 
becomes $Q$-normal, $[\hat{H}^{\dag^Q} , \hat{H} ] =0$, 
where the $Q$-Hermitian conjugate of any operator $A$, 
$A^{\dag^Q} \equiv Q^{-1} A^\dag Q$, is defined so that 
$\langle \psi_1 |_Q A | \psi_2 \rangle^* = \langle \psi_2 |_Q A^{\dag^Q} | \psi_1 \rangle$. 
Also, we define ${\dag^Q}$ for kets and bras by 
$| \psi_1 \rangle^{\dag^Q} \equiv \langle \psi_1 |_Q$, 
$\left(\langle \psi_1 |_Q \right)^{\dag^Q} \equiv | \psi_1 \rangle$. 
We argued that in the case of non-normal Hamiltonians we had better readjust 
the Hilbert space inner product, which will have 
a physical significance by delivering a Born rule of probabilities 
to the properly modified one defined by Eqs.(\ref{Qin}) and (\ref{Qin2}) 
so that unphysical transitions 
between energy eigenstates $|\lambda_i \rangle_1$ and $| \lambda_j \rangle_1$ 
with different eigenvalues are prohibited, i.e. not observed 
with an energy-conserving measurement instrument.

It is natural to attempt to choose $Q$ as close to the unit operator 
as possible to change the inner product in the Hilbert space as little as possible. 
In Refs.\cite{Nagao:2010xu,Nagao:2011za,Nagao:2012mj} 
we chose
\begin{equation}
Q= (P^\dag)^{-1} P^{-1} , \label{Q_Pdag-1P-1}
\end{equation}
where $P= ( | \lambda_1 \rangle_1, | \lambda_2 \rangle_1, \ldots )$ 
is a diagonalizing operator of $\hat{H}$, $\hat{H}=PD P^{-1}$. 
Incidentally, $P^{-1}$ is expressed as 
\begin{equation}
P^{-1} = \left( 
\begin{array}{c}
      {}_2\langle \lambda_1 |   \\
      {}_2\langle \lambda_2 |   \\
      \vdots   
\end{array}
\right) ,  
\end{equation} 
where the $| \lambda_j \rangle_2$ are the eigenstates of $\hat{H}^\dag$, 
\begin{equation}
| \lambda_j \rangle_2 = Q | \lambda_j \rangle_1 . \label{lambda2stateQlambda1state}
\end{equation}
We introduce an orthonormal basis $| e_i \rangle ~(i=1, \ldots)$ satisfying 
$\langle e_i | e_j \rangle = \delta_{ij}$ by 
$D | e_i \rangle = \lambda_i   | e_i \rangle$. 
Then, $P$, which obeys $| \lambda_i \rangle_1 = P | e_i \rangle$, is rewritten as 
$P=\sum_i | \lambda_i \rangle_1 ~\langle e_i |$, 
and $Q$ given in Eq.(\ref{Q_Pdag-1P-1}) is expressed as 
\begin{equation}
Q=\left( \sum_i | \lambda_i \rangle_1 ~{}_1\langle \lambda_i | \right)^{-1} 
= \sum_i | \lambda_i \rangle_2 ~{}_2\langle \lambda_i | . \label{form_Q_eigenstates}
\end{equation}
The completeness relation is written as 
$\sum_i | \lambda_i \rangle_1 ~{}_1 \langle \lambda_i |_Q 
=\sum_i | \lambda_i \rangle_2 ~{}_2 \langle \lambda_i |_{Q^{-1}} = 1$.

We note that the operator $Q$ is not unambiguously determined by 
the defining properties of Eqs.(\ref{Qin}) and (\ref{Qin2}),  
because if we define a Hermitian operator $Q_g$ by using some function of 
the Hamiltonian operator $g(\hat{H})$ by 
\begin{equation}
Q = \left\{ g(\hat{H}) \right\}^\dag Q_g g(\hat{H}) , \label{ambiguity0} 
\end{equation}
then Eq.(\ref{Qin2}) is rewritten as 
${}_1^g \langle \lambda_i | Q_g |  \lambda_j \rangle_1^g  
= \delta_{ij}$, 
where $|  \lambda_i \rangle_1^g$ is defined by 
$|  \lambda_i \rangle_1^g \equiv g(\hat{H}) | \lambda_i \rangle_1$. 
If, however, we write conditions involving $Q$ 
and operators not commuting 
with $\hat{H}$, such conditions will specify how to resolve 
the ambiguity by Eq.(\ref{ambiguity0}).

\subsection{Choice of $Q$ in the harmonic oscillator model}\label{choice_Q}

In the harmonic oscillator model, Eq.(\ref{lambda2stateQlambda1state}) is expressed as  
\begin{equation}
| n \rangle_2 = Q | n \rangle_1 
\quad
\Leftrightarrow
\quad 
{}_2 \langle n |  = {}_1 \langle n |_Q ,  \label{Qdef}
\end{equation}
and Eq.(\ref{form_Q_eigenstates}) provides the expression for $Q$: 
\begin{equation}
Q=\left( \sum_n | n \rangle_1 ~{}_1\langle n | \right)^{-1} 
= \sum_n | n \rangle_2 ~{}_2\langle n | . \label{expression_Q_eigenstates_ho}
\end{equation}
We investigate the properties of 
the operators $\hat{a}_1^{\dag^Q}$ and 
$\hat{a}_2^{\dag^{Q^{-1}}}$ expressed as 
\begin{eqnarray}
&&\hat{a}_1^{\dag^Q} = Q^{-1} \hat{a}_1^\dag Q , \label{a1dagQa1}     \\
&&\hat{a}_2^{\dag^{Q^{-1}}} = Q \hat{a}_2^\dag Q^{-1} .  \label{a2dagQa2}
\end{eqnarray}
The operators $\hat{a}_1^{\dag^Q}$ and $\hat{a}_2^{\dag^{Q^{-1}}}$ obey 
\begin{eqnarray}
&&\hat{a}_1^{\dag^Q} | n \rangle_1 
= \sqrt{n+1} | n +1 \rangle_1 ,  \\
&&\hat{a}_2^{\dag^{Q^{-1}}} | n \rangle_2 
= \sqrt{n+1} | n +1 \rangle_2 ,  
\end{eqnarray}
where we have used Eqs.(\ref{a1dagstep}) and (\ref{a2dagstep}), respectively. 
Comparing these relations with Eqs.(\ref{a2dagstep}) and (\ref{a1dagstep}),  and 
using Eq.(\ref{expression_Q_eigenstates_ho}), we obtain the following relations: 
\begin{eqnarray}
&&\hat{a}_1^{\dag^Q} = \hat{a}_2^\dag = \sum_{n=0} \sqrt{n+1} | n+1 \rangle_1 ~{}_2 \langle n | , 
\label{choiceQinho} \\
&&\hat{a}_2^{\dag^{Q^{-1}}} = \hat{a}_1^\dag  = \sum_{n=0} \sqrt{n+1} | n+1 \rangle_2 ~{}_1 \langle n |  . \label{choiceQinho2}
\end{eqnarray}
Equation (\ref{choiceQinho2}) is also provided by operating $Q$ and $Q^{-1}$ 
from the left and right respectively on both sides of Eq.(\ref{choiceQinho}). 
Using Eqs.(\ref{choiceQinho}), (\ref{choiceQinho2}), (\ref{newqhat}), and (\ref{newphat}), 
we obtain the relations  
\begin{eqnarray}
&&\hat{q}_\mathrm{new}^{\dag_Q} = Q^{-1} \hat{q}_\mathrm{new}^\dag Q = e^{i\theta} \hat{q}_\mathrm{new}, \label{Q-1qhatnewdagQ}  \\
&&\hat{p}_\mathrm{new}^{\dag_Q} = Q^{-1} \hat{p}_\mathrm{new}^\dag Q = e^{-i\theta} \hat{p}_\mathrm{new}, \label{Q-1phatnewdagQ}  
\end{eqnarray}
where $\theta = \arg(m \omega)$ was introduced in Eq.(\ref{momega}). 
We note that Eq.(\ref{choiceQinho}) or the pair of 
Eqs.(\ref{Q-1qhatnewdagQ}) and (\ref{Q-1phatnewdagQ}) can be regarded as 
conditions that $Q$ has to obey.  
Indeed, they can determine $Q$ up to an overall factor. 
In our present construction, $Q$ is defined by 
Eqs.(\ref{expression_Q_eigenstates_ho}) and (\ref{HPS}), 
so $Q$, whose overall factor is already determined, 
obeys Eqs.(\ref{choiceQinho}), (\ref{Q-1qhatnewdagQ}), and (\ref{Q-1phatnewdagQ}) 
automatically.

Using Eqs.(\ref{choiceQinho}) and (\ref{choiceQinho2}), we can rewrite 
the number operators defined in Eqs.(\ref{nhat1}) and (\ref{nhat2}) 
in more usual expressions as  
$\hat{n}_1 = \hat{a}_1^{\dag^Q} \hat{a}_1$ and $\hat{n}_2 = \hat{a}_2^{\dag^{Q^{-1}}} \hat{a}_2$, 
which are $Q$-Hermitian and $Q^{-1}$-Hermitian respectively, 
and $\hat{H}$ and $\hat{H}^\dag$ given in Eqs.(\ref{H_hbaromega_n}) and (\ref{Hdagger_hbaromega*_n}) 
as 
$\hat{H}= \hbar \omega  \left(\hat{a}_1^{\dag^Q} \hat{a}_1 + \frac{1}{2} \right)$ and 
$\hat{H}^\dag =\hbar \omega^* \left( \hat{a}_2^{\dag^{Q^{-1}}} \hat{a}_2  + \frac{1}{2} \right)$. 
Since $\hat{H}^{\dag^Q}$ is written as 
\begin{eqnarray}
\hat{H}^{\dag^Q} 
&=& 
\hbar \omega^*  \left(\hat{n}_1 + \frac{1}{2} \right)  
= \frac{\omega^*}{\omega} \hat{H} , \label{HdagQ}
\end{eqnarray}
$\hat{H}$ only deviates from $Q$-Hermiticity because of $\omega$ 
being complex.

Using the inner product $I_Q$ instead of the usual inner product 
in the Hilbert space, 
we have achieved a formalism that is very similar to the usual one in the RAT. 
We defined $\hat{a}_1$ and $\hat{a}_1^{\dag^Q} = \hat{a}_2^\dag$ 
as annihilation and creation operators respectively for $| n \rangle_1$, 
and $\hat{a}_2$ and $\hat{a}_2^{\dag^{Q^{-1}}} = \hat{a}_1^\dag$ for $| n \rangle_2$. 
Our $|n \rangle_1$ is ``$Q$-orthonormal", i.e. orthonormal with respect to the 
inner product $I_Q$, while $|n \rangle_2$ is  ``$Q^{-1}$-orthonormal". 
Indeed, using Eq.(\ref{Qdef}), we can rewrite Eq.(\ref{2branmket1}) as 
\begin{equation} 
{}_1\langle m |_Q n \rangle_1 
={}_2\langle m |_{Q^{-1}} n \rangle_2
= \delta_{m n}.  \label{Qnormalization}
\end{equation}
Thus, the dual normalization of Eq.(\ref{2branmket1}) can be interpreted as 
``$Q$-normalization'' for $|n \rangle_1$ or ``$Q^{-1}$-normalization" for $|n \rangle_2$, 
as expressed by Eq.(\ref{Qnormalization}).

\section{The maximization principle and the solution to the harmonic oscillator model} 
\label{sec:max_prin_solution_to_ho}

In the future-included CAT,  we suppose that 
$| A(T_A) \rangle$ and $|B(T_B) \rangle$ are randomly given at first, i.e. 
they are given by the overlaps of many states. 
However, due to the existence of the imaginary part of the action $S_\mathrm{I}$, only a single class 
of pairs of $| A(t) \rangle$ and $|B(t) \rangle$ dominates most significantly in the FPI. 
Then we can approximate $| A(t) \rangle$ and $|B(t) \rangle$ by such representative 
states, and classical physics is described by them.  
Indeed, in Refs.~\cite{Nagao:2015bya, Nagao:2017cpl, Nagao:2017book, Nagao:2017ztx} 
we argued by such a maximization principle that we can obtain real expectation values. 
In the RAT, classical behaviors are typically described by coherent states, so 
it would be natural for us to expect that coherent states work similarly even in the CAT.  
Supposing that we utilize the maximization principle, we can imagine a simple 
situation where the representative $| A(t) \rangle$ and $|B(t) \rangle$ are essentially approximated 
by just a pair of coherent states. 
In this section, based on this speculation, 
we first consider such a simple situation where $| A(t) \rangle$ and $|B(t) \rangle$ are 
given by a single pair of coherent states as a preliminary study. 
Supposing that they time-develop according to the Schr\"{o}dinger equations, 
we see that we can obtain an equation of motion. 
Next, briefly explaining the maximization 
principle~\cite{Nagao:2015bya, Nagao:2017cpl, Nagao:2017book, Nagao:2017ztx}, 
and applying it to the harmonic oscillator model, we argue that 
the system obtained is described by a $Q$-Hermitian Hamiltonian, which 
can be expressed in terms of $Q$-Hermitian coordinate and momentum operators. 
Finally, we find that the generic solution to the harmonic oscillator model is the ground state.

In the following, we adopt the proper inner product $I_Q$ 
for all quantities. 
This is realized by changing the notation of  the final state $\langle B(T_B) |$ 
as $\langle B(T_B) | \rightarrow \langle B(T_B) |_Q$. 
Then $\langle B(T_B) |$ time-develops not according to 
Eq.(\ref{schro_eq_Bstate_old}) but to 
\begin{eqnarray}
-i \hbar \frac{d}{dt} \langle B(t) |_Q  
= \langle B(t) |_Q  \hat{H} 
\quad \Leftrightarrow \quad 
i \hbar \frac{d}{dt} | B(t) \rangle = {\hat{H}}^{\dag^Q} | B(t) \rangle ,  
\label{schro_eq_Bstate} 
\end{eqnarray}
and the normalized matrix element $\langle \hat{\cal O} \rangle^{BA}$ in Eq.(\ref{OBA}), 
which is a strong candidate for the expectation value  of the operator $\hat{\cal O}$, 
is replaced with  
\begin{equation}
\langle \hat{\cal O} \rangle_Q^{BA} 
\equiv \frac{ \langle B(t) |_Q  \hat{\cal O}  | A(t) \rangle }{ \langle B(t) |_Q A(t) \rangle }.  
\label{OQBA}
\end{equation} 
In addition, we suppose that $|A(T_A) \rangle$ and $|B(T_B) \rangle$ are 
$Q$-normalized, i.e. normalized with the modified inner product $I_Q$, by 
$\langle A(T_A) |_{Q} A(T_A) \rangle =1$ and $\langle B(T_B) |_{Q} B(T_B) \rangle = 1$, 
respectively.

\subsection{Preliminary study in the case of $| A(T_A) \rangle$ and $| B(T_B) \rangle$ being  coherent states} \label{subsec:ABstates_being_coherent_states}

As a preliminary study, 
based on the speculation that classical behaviors are typically 
described by coherent states even in the CAT, 
let us  consider a situation where $| A(t) \rangle$ and $|B(t) \rangle$ are 
given by a pair of coherent states 
$|\lambda_A (t)  \rangle_{\mathrm{coh}, 1}$ and $|\lambda_B (t)  \rangle_{\mathrm{coh}, 1}$, 
which are defined in Eqs.(\ref{lambdaket1}) and (\ref{fn}), 
and investigate  how $\langle \hat{\cal O} \rangle_Q^{BA}$ behaves. 
To study this, let us formulate the time-development of the coherent states.

\subsubsection{Time-development of coherent states}

We consider the case where  $| A(T_A) \rangle$ and $| B(T_B) \rangle$  are given by 
the coherent states $|\lambda_A (T_A)  \rangle_{\mathrm{coh}, 1}$ and $|\lambda_B (T_B)  \rangle_{\mathrm{coh}, 1}$ 
that time-develop according to the Schr\"{o}dinger equations 
\begin{eqnarray}
&&i\hbar \frac{d}{dt}|\lambda_A (t)  \rangle_{\mathrm{coh}, 1}
= H |\lambda_A (t)  \rangle_{\mathrm{coh}, 1} , \label{Sch-coh-1}  \\
&&i\hbar \frac{d}{dt}|\lambda_B (t)  \rangle_{\mathrm{coh}, 1}
= H^{\dag^Q} |\lambda_B (t)  \rangle_{\mathrm{coh}, 1} ,  \label{Sch-coh-2}
\end{eqnarray}
and are normalized with the modified inner product $I_Q$ by 
${}_{\mathrm{coh},1} \langle \lambda_A (T_A) |_Q \lambda_A (T_A)  \rangle_{\mathrm{coh}, 1} =1$ 
and 
${}_{\mathrm{coh},1}\langle \lambda_B (T_B) |_Q \lambda_B (T_B)  \rangle_{\mathrm{coh}, 1} =1$,   
respectively. 
Then 
$|\lambda_A (t)  \rangle_{\mathrm{coh}, 1}$ and $|\lambda_B (t)  \rangle_{\mathrm{coh}, 1}$ 
are expressed as 
\begin{eqnarray}
|\lambda_A (t)  \rangle_{\mathrm{coh}, 1}
&=& 
e^{-i\frac{\omega}{2}(t-T_A)}
e^{-\frac{ | \lambda_A (T_A) |^2}{2}} 
\exp[ \lambda_A (T_A) e^{-i\omega(t-T_A)} \hat{a}_2^\dag ] 
| 0 \rangle_1 \nonumber \\ 
&=& 
e^{-i\frac{\omega}{2}(t-T_A)}
\exp\left( 
- \frac{ | \lambda_A (T_A) |^2}{2} \left\{ 1- \exp \left[ 2 \omega_I (t-T_A) \right] \right\} 
\right)  
| \lambda_A (T_A) e^{-i\omega(t-T_A)} \rangle_{\mathrm{coh}, 1} , \nonumber \\ 
\label{coh-1-t-sch-n-TA} \\
|\lambda_B (t)  \rangle_{\mathrm{coh}, 1}
&=& 
e^{-i\frac{\omega^*}{2}(t-T_B)}
e^{-\frac{ | \lambda_B (T_B) |^2}{2}}  
\exp[ \lambda_B (T_B) e^{-i\omega^* (t-T_B)} \hat{a}_2^\dag ] 
| 0 \rangle_1 \nonumber \\ 
&=& 
e^{-i\frac{\omega^*}{2}(t-T_B)} 
\exp\left( 
- \frac{ | \lambda_B (T_B) |^2}{2} \left\{ 1- \exp \left[ - 2 \omega_I (t-T_B) \right] \right\} 
\right) 
| \lambda_B (T_B) e^{-i\omega^*(t-T_B)} \rangle_{\mathrm{coh}, 1} . \nonumber \\ 
\label{coh-2-t-sch-n-TB} 
\end{eqnarray}
Operating $\hat{a}_1$ on both sides of Eqs.(\ref{coh-1-t-sch-n-TA}) and.(\ref{coh-2-t-sch-n-TB}), 
we obtain the relations 
\begin{eqnarray}
&&\hat{a}_1 |\lambda_A (t)  \rangle_{\mathrm{coh}, 1}
=
\lambda_A (T_A) e^{-i\omega(t-T_A)} |\lambda_A (t)  \rangle_{\mathrm{coh}, 1} , 
\label{alambdaA(t)coh-1-t-sch-n-TA} \\
&&\hat{a}_1 |\lambda_B (t)  \rangle_{\mathrm{coh}, 1} 
=
\lambda_B (T_B) e^{-i\omega^*(t-T_B)} |\lambda_B (t)  \rangle_{\mathrm{coh}, 1} , 
\label{alambdaB(t)coh-2-t-sch-n-TB}
\end{eqnarray}
where 
we have used Eqs.(\ref{a1lambdaketcoh1}), (\ref{coh-1-t-sch-n-TA}), and (\ref{coh-2-t-sch-n-TB}). 
Equations (\ref{alambdaA(t)coh-1-t-sch-n-TA}) and (\ref{alambdaB(t)coh-2-t-sch-n-TB}) 
suggest that $\lambda_A(t)$ and $\lambda_B(t)$ time-develop as 
\begin{eqnarray}
&&\lambda_A(t) = \lambda_A (T_A) e^{-i\omega(t-T_A)} ,  \label{lambdaA(t)lambdaA(TA)} \\
&& \lambda_B(t) = \lambda_B (T_B) e^{-i\omega^*(t-T_B)} ,  \label{lambdaB(t)lambdaB(TB)}
\end{eqnarray}
so that we have relations similar to Eq.(\ref{a1lambdaketcoh1}): 
\begin{eqnarray}
&&\hat{a}_1 |\lambda_A (t)  \rangle_{\mathrm{coh}, 1} =
\lambda_A (t) 
|\lambda_A (t)  \rangle_{\mathrm{coh}, 1} , \label{a1_lambdaAtcohstate=lambdaAt_lambdaAtcohstate} \\
&&\hat{a}_1 |\lambda_B (t)  \rangle_{\mathrm{coh}, 1} =
\lambda_B (t) 
|\lambda_B (t)  \rangle_{\mathrm{coh}, 1} . \label{a1_lambdaBtcohstate=lambdaBt_lambdaBtcohstate}
\end{eqnarray}
%

\subsubsection{Derivation of classical equation of motion}

Now we are prepared for evaluating 
$\langle \hat{q}_\mathrm{new} \rangle_Q^{\lambda_B \lambda_A}$ 
and $\langle \hat{p}_\mathrm{new} \rangle_Q^{\lambda_B \lambda_A}$, 
where $\langle \hat{\cal O} \rangle_Q^{BA}$ for any operator $\hat{\cal O}$ 
is defined in Eq.(\ref{OQBA}). 
They are calculated as 
\begin{eqnarray}
\langle \hat{q}_\mathrm{new}  \rangle_Q^{\lambda_B \lambda_A} 
&=& \sqrt{ \frac{\hbar}{ 2 m \omega} } ( \lambda_A (t) + \lambda_B (t)^* ) , 
\label{exp_value_q_BA} \\
\langle  \hat{p}_\mathrm{new} \rangle_Q^{\lambda_B \lambda_A} 
&=& -i \sqrt{ \frac{\hbar m \omega}{2}}  ( \lambda_A (t) - \lambda_B (t)^* ) , 
\label{exp_value_p_BA}  
\end{eqnarray} 
where we have used 
Eqs.(\ref{newqhat}), (\ref{newphat}), (\ref{choiceQinho}),  (\ref{a1_lambdaAtcohstate=lambdaAt_lambdaAtcohstate}), and  (\ref{a1_lambdaBtcohstate=lambdaBt_lambdaBtcohstate}). 
Equations (\ref{lambdaA(t)lambdaA(TA)}) and (\ref{lambdaB(t)lambdaB(TB)}) 
suggest that 
$\dot{\lambda_B}(t)$ and $\dot{\lambda_A}(t)$ are expressed as 
$\dot{\lambda_B}(t)= -i\omega^* \lambda_B(t)$ and $\dot{\lambda_A}(t)= -i\omega \lambda_A(t)$. 
Using these relations, we can evaluate the time derivative of 
Eqs.(\ref{exp_value_q_BA}) and (\ref{exp_value_p_BA}) as follows:  
\begin{eqnarray}
\frac{d}{dt}  \langle \hat{q}_\mathrm{new}  \rangle_Q^{\lambda_B \lambda_A}     
&=& \frac{1}{m}   \langle  \hat{p}_\mathrm{new} \rangle_Q^{\lambda_B \lambda_A}  , \label{ddtqhatBA} \\
\frac{d}{dt} \langle  \hat{p}_\mathrm{new} \rangle_Q^{\lambda_B \lambda_A}  
&=& - m \omega^2 \langle \hat{q}_\mathrm{new}  \rangle_Q^{\lambda_B \lambda_A}   
= - \langle V'(\hat{q}_\mathrm{new})  \rangle_Q^{\lambda_B \lambda_A} ,
\label{ddtphatBA}
\end{eqnarray}
where $V$ is the potential of the harmonic oscillator, which is given in Eq.(\ref{ho_potential_hat}). 
Equations (\ref{ddtqhatBA}) and (\ref{ddtphatBA}) are 
the momentum relation and equation of motion, which are 
consistent with Eqs.(\ref{dqcdt}) and (\ref{dpcdt}).  
As when we reviewed the general properties of the future-included theory~\cite{Nagao:2012mj} 
in Sect.~\ref{sec:review_future-included_theory},  
we have obtained Ehrenfest's theorem: 
$m\frac{d^2}{dt^2} \langle \hat{q}_\mathrm{new} \rangle_Q^{\lambda_B \lambda_A} 
= - \langle V'(\hat{q}_\mathrm{new}) \rangle_Q^{\lambda_B \lambda_A}$,  
and $\langle {\cal O} \rangle_Q^{\lambda_B \lambda_A}$ provides the saddle point development with $t$. 
It is very nice to have such properties. 
Though $\langle {\cal O} \rangle_Q^{\lambda_B \lambda_A}$ is generically complex, 
if a pair of coherent states with $\lambda_A (t)$ and $\lambda_B (t)$ such that 
$\langle {\cal O} \rangle_Q^{\lambda_B \lambda_A}$ becomes real dominates most significantly 
in the FPI, 
then classical physics is nicely realized. 
In the next subsection, 
to solve the harmonic oscillator model 
we utilize the maximization principle and investigate what kind of 
$| A(t) \rangle$ and $ | B(t) \rangle$ dominate most significantly in the FPI. 
We shall find that they are not such interesting coherent states, but just the ground state.

\subsection{Application of the maximization principle to the harmonic oscillator model}
\label{sec:application_of_maximization_principle}

First we explain the maximization principle briefly. 

\vspace{0.5cm}

\noindent
{\bf Theorem 1. Maximization principle in the future-included theories} \\ 
{\em 
As a prerequisite, assume that a given Hamiltonian 
$\hat{H}$ is non-normal but diagonalizable 
and that the imaginary parts of the eigenvalues 
of $\hat{H}$ are bounded from above; then 
define a modified inner product $I_Q$ by means 
of a Hermitian operator $Q$ arranged so 
that $\hat{H}$ becomes normal with respect to $I_Q$. 
Let the two states $| A(t) \rangle$ and $ | B(t) \rangle$ 
time-develop according to the Schr\"{o}dinger equations 
with $\hat{H}$ and $\hat{H}^{\dag^Q}$ respectively: 
$|A (t) \rangle = e^{-\frac{i}{\hbar}\hat{H} (t-T_A) }| A(T_A) \rangle$, 
$|B (t) \rangle = e^{-\frac{i}{\hbar} {\hat{H}}^{\dag^Q} (t-T_B) } | B(T_B)\rangle$, 
and be normalized with $I_Q$ 
at the initial time $T_A$ and the final time $T_B$ respectively: 
$\langle A(T_A) |_{Q} A(T_A) \rangle = 1$,
$\langle B(T_B) |_{Q} B(T_B) \rangle = 1$. 
Next, determine $|A(T_A) \rangle$ and $|B(T_B) \rangle$ so as to maximize 
the absolute value of the transition amplitude 
$|\langle B(t) |_Q A(t) \rangle|=|\langle B(T_B)|_Q \exp(-i\hat{H}(T_B-T_A)) |A(T_A) \rangle|$. 
Then, provided that an operator $\hat{\cal O}$ is $Q$-Hermitian, i.e. Hermitian 
with respect to the inner product $I_Q$, 
$\hat{\cal O}^{\dag^Q} = \hat{\cal O}$, 
the normalized matrix element of the operator $\hat{\cal O}$ defined by 
$\langle \hat{\cal O} \rangle_Q^{BA} 
\equiv
\frac{\langle B(t) |_Q \hat{\cal O} | A(t) \rangle}{\langle B(t) |_Q A(t) \rangle}$ 
becomes {\rm real} and time-develops under 
a {\rm $Q$-Hermitian} Hamiltonian. }


\vspace*{0.5cm}


In this theorem\footnote{For a normal Hamiltonian $\hat{H}$, 
the above theorem becomes simpler with $Q=1$.}, exactly speaking, not only the maximizing states but also many other states contribute to the transition amplitude, but their contribution becomes very small 
for large $T=T_B-T_A$, in which we are interested practically. 
So, we ignore the effects of the other states, 
and consider only those of the maximizing states. 
Then, the normalized matrix element $\langle \hat{\cal O} \rangle_Q^{BA}$ 
for a $Q$-Hermitian operator $\hat{\cal O}$ turns out to be real, 
and time-develops according to a $Q$-Hermitian Hamiltonian. 
We call this way of thinking the maximization principle. 
This theorem can be applied not only to the CAT but also to the RAT. 
In the CAT  there are imaginary parts of the eigenvalues of $\hat{H}$, 
$\text{Im}\lambda_i$, and the eigenstates having the largest 
$\text{Im}\lambda_i$ blow up and contribute most to the the absolute value 
of the transition amplitude $|\langle B(t) |_Q A(t) \rangle|$. 
Utilizing this property, we proved the theorem in the case of the CAT~\cite{Nagao:2015bya}. 
On the other hand, in the RAT, there are no $\text{Im}\lambda_i$, 
so the full set of eigenstates of $\hat{H}$ can contribute to 
$| \langle B (t) | A (t) \rangle |$~\cite{Nagao:2017cpl}. 
The theorem is reviewed in Refs.~\cite{Nagao:2017book, Nagao:2017ztx}.

Now we try to apply the maximization principle to the harmonic oscillator model. 
$| A (T_A) \rangle$ and $| B (T_B) \rangle$ time-develop as Eqs.(\ref{schro_eq_Astate})
and (\ref{schro_eq_Bstate}), and are $Q$-normalized by 
$\langle A(T_A) |_{Q} A(T_A) \rangle =1$ and $\langle B(T_B) |_{Q} B(T_B) \rangle = 1$. 
The normalized matrix element $\langle \hat{\cal O} \rangle_Q^{BA}$ is given in Eq.(\ref{OQBA}). 
In addition, in the harmonic oscillator model the eigenvalue of the Hamiltonian 
for $| n \rangle_1$, $\lambda_n$, is given in Eq.(\ref{lambda_n}). 
So $\text{Re}\lambda_n = \hbar \text{Re}\omega \left( n+ \frac{1}{2} \right)$ 
and $\text{Im}\lambda_n= \hbar \text{Im}\omega \left( n+ \frac{1}{2} \right)$. 
To consider the theorem explicitly,  
let us expand $| A(t) \rangle$ and $| B(t) \rangle$ in terms of the eigenstates $| n \rangle_1$ 
as follows: 
\begin{eqnarray}
&&|A (t) \rangle = \sum_n a_n (t) | n \rangle_1, \label{Aketexpansion}\\
&&|B (t) \rangle = \sum_n b_n (t) | n \rangle_1,  \label{Bketexpansion}
\end{eqnarray}
where $a_n (t)$ and $b_n (t)$ are expressed as 
\begin{eqnarray}
&&a_n (t) = a_n (T_A) e^{-i \omega \left( n+ \frac{1}{2} \right) (t-T_A) }, \label{aitimedevelopment} \\ 
&&b_n (t) = b_n (T_B) e^{-i \omega^* \left( n+ \frac{1}{2} \right) (t-T_B) }. \label{bitimedevelopment} 
\end{eqnarray}
We write  $a_n(T_A)$ and $b_n(T_B)$ as 
$a_n(T_A)= | a_n(T_A) | e^{i \theta_{a_n}}$ and $b_n(T_B) = | b_n(T_B) | e^{i \theta_{b_n}}$, 
and introduce 
\begin{eqnarray}
&&T\equiv T_B - T_A , \label{T} \\
&&\Theta_n \equiv \theta_{a_n} - \theta_{b_n} 
- T \text{Re}\omega \left( n+ \frac{1}{2} \right) , \label{Theta_n} \\
&&R_n \equiv |a_n (T_A)| |b_n (T_B)| 
e^{T \text{Im}\omega \left( n+ \frac{1}{2} \right) }. 
\end{eqnarray} 
Then, since $\langle B (t) |_Q A (t) \rangle$ is expressed as 
$\langle B (t) |_Q A (t) \rangle = \sum_n R_n e^{i \Theta_n}$, 
$| \langle B (t) |_Q A (t) \rangle |^2$ is calculated as 
\begin{equation}
| \langle B (t) |_Q A (t) \rangle |^2
= \sum_n R_n^2 + 2 \sum_{n<m} R_n R_m \cos(\Theta_i - \Theta_j). \label{|BbraQAket|^2}
\end{equation}
In addition, the normalization conditions for $| A(T_A) \rangle$ and $| B(T_B) \rangle$ 
are expressed as $\sum_n  | a_n (T_A) |^2  = \sum_n  | b_n (T_B) |^2  = 1$. 
We note that, since we are studying the harmonic oscillator model 
in the whole parallelogram region allowed by Eqs.(\ref{mIgeq02}) and (\ref{Immomega2leq02}) 
except for the two corners $(\theta_m, \theta_\omega)=(0, - \frac{\pi}{2}), (\pi, - \frac{\pi}{2})$ 
in the phase diagram given in Fig.~\ref{fig:phase_diagram_ho}, 
the imaginary part of the angular frequency $\omega$ is negative, $\text{Im} \omega \le 0$.

Let us first consider the case where $\text{Im} \omega < 0$. 
The imaginary parts of the 
eigenvalues of the Hamiltonian, $\text{Im} \lambda_n$, are supposed to be bounded from above 
to avoid the FPI $\int e^{\frac{i}{\hbar}S} {\cal D} \text{path}$ 
being divergently meaningless. 
So some of $\text{Im} \lambda_n$ take the maximal value $B$. 
We denote the corresponding subset of $\{ n \}$ as $A$. 
$\text{Im}\lambda_n  = \hbar \text{Im}\omega \left( n+ \frac{1}{2} \right)$ 
can take the maximum value $B= \frac{\hbar}{2} \text{Im}\omega$ only for $n=0$, for which 
$\text{Re}\lambda_0 = \frac{\hbar}{2} \text{Re}\omega$ 
and $\text{Im}\lambda_0 = \frac{\hbar}{2} \text{Im}\omega$. 
Hence we find that, in the harmonic oscillator model,  $A=\{ 0 \}$. 
Then, since $R_n \geq 0$, $| \langle B(t) |_Q A(t) \rangle |$ can take the maximal value 
$e^{\frac{1}{\hbar}TB}= e^{\frac{T}{2}\text{Im}\omega}$ 
only under the following conditions: 
\begin{eqnarray} 
&&  | a_0 (T_A) | = |b_0 (T_B)| = 1,  \label{nc_ATABTB3} \\
&& | a_n (T_A) |  = | b_n (T_B) | =0 \quad \text{for $\forall n$ s.t. $n \neq 0$} , \label{abinotinA0} 
\end{eqnarray}
and the states to maximize $| \langle B (t) |_Q A (t) \rangle |$, 
$| A(t) \rangle_{\rm{max}}$ and $| B(t) \rangle_{\rm{max}}$, are expressed as 
\begin{eqnarray}
&&|A (t) \rangle_{\rm{max}} = a_0 (t) | 0 \rangle_1 , 
\label{Atketmax_sum_inA_ai} \\
&&|B (t) \rangle_{\rm{max}} = b_0 (t) | 0 \rangle_1 , 
\label{Btketmax_sum_inA_bi} 
\end{eqnarray}
where $a_0 (t)$ and $b_0 (t)$ obey Eq.(\ref{nc_ATABTB3}).
That is to say, the ground state $| 0 \rangle_1$ is chosen for both 
the maximizing states $| A(t) \rangle_{\rm{max}}$ and $| B(t) \rangle_{\rm{max}}$.

To evaluate $\langle \hat{\cal O} \rangle_Q^{BA}$ 
for $| A(t) \rangle_{\rm{max}} $ and $| B(t) \rangle_{\rm{max}}$,  
utilizing the $Q$-Hermitian part of $\hat{H}$, 
$\hat{H}_{Qh} \equiv \frac{1}{2} \left( \hat{H} +  \hat{H}^{\dag^Q}  \right)$, 
we define the following state: 
\begin{equation}
| \tilde{A}(t) \rangle \equiv 
e^{-\frac{i}{\hbar}(t-T_A) \hat{H}_{Qh}} | A(T_A) \rangle_{\rm{max}}, 
\end{equation}
which is normalized as $\langle \tilde{A}(t) |_Q \tilde{A}(t) \rangle = 1$ 
and obeys the Schr\"{o}dinger equation 
\begin{eqnarray}
i\hbar  \frac{d}{d t}| \tilde{A}(t) \rangle 
&=& \hat{H}_{Qh} | \tilde{A}(t) \rangle .  \label{ScheqAtildetket}
\end{eqnarray} 
Using Eqs.(\ref{nc_ATABTB3}) and (\ref{abinotinA0}), we obtain 
${}_{\rm{max}} \langle B (t) |_Q A (t) \rangle_{\rm{max}} 
=e^{i \Theta_0} R_0  = e^{i \Theta_0} e^{\frac{B T}{\hbar} }$, and 
\begin{eqnarray} 
{}_{\rm{max}} \langle B (t) |_Q \hat{\cal O} | A  (t) \rangle_{\rm{max}} 
&=& 
e^{i \Theta_0} e^{\frac{B T}{\hbar} }
\langle \tilde{A}(t) |_Q \hat{\cal O} | \tilde{A}(t) \rangle \nonumber \\
&=& 
e^{i \Theta_0} e^{\frac{B T}{\hbar} }
a_0(T_A)^*  a_0(T_A)  
{}_1\langle 0 |_Q \hat{\cal O} | 0 \rangle_1  \nonumber \\
&=& 
e^{i \Theta_0} e^{\frac{B T}{\hbar} }
{}_{\rm{max}} \langle A(T_A) |_Q \hat{\cal O} | A (T_A) \rangle_{\rm{max}}. 
\end{eqnarray}
Thus, $\langle \hat{\cal O} \rangle_Q^{BA}$  
for $| A(t) \rangle_{\rm{max}}$ and $| B(t) \rangle_{\rm{max}}$ 
is evaluated as 
\begin{eqnarray}
\langle \hat{\cal O} \rangle_Q^{B_{\rm{max}} A_{\rm{max}}} 
&=& 
\langle \tilde{A}(t) |_Q \hat{\cal O} | \tilde{A}(t) \rangle 
\equiv 
\langle \hat{\cal O} \rangle_Q^{\tilde{A} \tilde{A}} . \label{OBAmaxtilde}
\end{eqnarray}
Since 
$\left\{ \langle \hat{\cal O} \rangle_Q^{\tilde{A} \tilde{A}} \right\}^*=\langle \hat{\cal O}^{\dag^Q} \rangle_Q^{\tilde{A} \tilde{A}}$, 
$\langle \hat{\cal O} \rangle_Q^{BA}$ 
for $| A(t) \rangle_{\rm{max}} $ and $| B(t) \rangle_{\rm{max}}$  
is real for $Q$-Hermitian $\hat{\cal O}$. 
In addition, if we express $\langle \hat{\cal O} \rangle_Q^{\tilde{A} \tilde{A}}$ as 
$\langle \hat{\cal O} \rangle_Q^{\tilde{A} \tilde{A}}
=\langle \tilde{A}(T_A) |_Q \hat{\cal O}_{H}(t, T_A) | \tilde{A}(T_A) \rangle$, 
where 
$\hat{\cal O}_{H}(t, T_A) 
\equiv 
e^{ \frac{i}{\hbar} \hat{H}_{Qh} (t-T_A) } 
\hat{\cal O} 
e^{ -\frac{i}{\hbar} \hat{H}_{Qh} (t-T_A)}$ is the Heisenberg operator, 
$\hat{\cal O}_{H}(t, T_A)$ obeys 
the Heisenberg equation 
$i\hbar  \frac{d}{d t} \hat{\cal O}_{H}(t, T_A) 
= [ \hat{\cal O}_{H}(t, T_A) , \hat{H}_{Qh} ]$, 
so 
$\langle \hat{\cal O} \rangle_Q^{\tilde{A} \tilde{A}}$ time-develops 
under the $Q$-Hermitian Hamiltonian $\hat{H}_{Qh}$ as 
\begin{eqnarray}
\frac{d}{dt} \langle \hat{\cal O} \rangle_Q^{\tilde{A} \tilde{A}} 
&=&
\frac{i}{\hbar} 
\langle \left[ \hat{H}_{Qh}, \hat{\cal O} \right]  
\rangle_Q^{\tilde{A} \tilde{A}} . 
\label{ddtOAtildeAtildeQ}
\end{eqnarray}
Thus the maximization principle generically provides both 
the reality of $\langle \hat{\cal O} \rangle_Q^{BA}$ 
for $Q$-Hermitian $\hat{\cal O}$ and the $Q$-Hermitian Hamiltonian $\hat{H}_{Qh}$.  
However, in the harmonic oscillator model that we are now studying 
we have the particular relation 
$\langle \hat{\cal O} \rangle_Q^{\tilde{A} \tilde{A}} 
={}_{\rm{max}} \langle A(T_A) |_Q \hat{\cal O} | A (T_A) \rangle_{\rm{max}}$, 
so $\langle \hat{\cal O} \rangle_Q^{\tilde{A} \tilde{A}}$ is constant in time: 
$\frac{d}{dt} \langle \hat{\cal O} \rangle_Q^{\tilde{A} \tilde{A}} =0$.

In the case where $\text{Im} \omega =0$ 
we are left only at the two corners $(\theta_m, \theta_\omega)=(0, 0), (\pi, -\pi)$ 
in the phase diagram shown in Fig.\ref{fig:phase_diagram_ho}, 
because the conditions in Eqs.(\ref{mIgeq02}) and (\ref{Immomega2leq02}) are imposed on 
$\theta_m$ and $\theta_\omega$. 
Since for $\forall n$ $\text{Im} \lambda_n = 0$, i.e. $\lambda_n \in \mathbf{R}$\footnote{Though 
both $m$ and $\omega$ are real, 
$\hat{H}$ is not Hermitian, $\hat{H}^\dag \neq \hat{H}$, because $\hat{H}$ 
includes $\hat{q}_\mathrm{new}$ and $\hat{p}_\mathrm{new}$. 
We might thus feel that we have encountered a contradiction, 
but this is not the case. 
We can circumvent this seeming contradiction by noticing that $\hat{H}$ is $Q$-Hermitian. }, 
the norms of $| A(t) \rangle$ and $| B(t) \rangle$ are constant in time:  
$\langle A(t) |_Q A(t) \rangle =\langle A(T_A) |_Q A(T_A) \rangle =1$, 
$\langle B(t) |_Q B(t) \rangle =\langle B(T_B) |_Q B(T_B) \rangle =1$.  
Therefore, we easily find that $| B(t) \rangle_{\rm{max}} = e^{-i\Theta_c} | A(t) \rangle_{\rm{max}}$, 
where $\Theta_c$ is a constant phase factor such that, 
for $\Theta_n$ given in Eq.(\ref{Theta_n}), $\Theta_n = \Theta_c$ for $\forall n$. 
Thus, in this special case $| A(t) \rangle_{\rm{max}}$ and $| B(t) \rangle_{\rm{max}}$ 
are not restricted to a unique pair of states. This is in contrast to the case 
where $\text{Im} \omega < 0$. 
Indeed, in the case where $\text{Im} \omega =0$ we have harmonic oscillators defined 
with real coefficients $m$ and $\omega$ as in the RAT\footnote{In the case 
where $\text{Im} \omega =0$, if we choose 
the Hamiltonian $\hat{H}_{\epsilon=\epsilon'=0}$ given in Eq.(\ref{H_epsilon=0}) 
on behalf of Eqs.(\ref{hoHamiltonian}) and (\ref{ho_potential_hat}) at the beginning, 
then harmonic oscillators become quite usual ones with $Q=1$ in the RAT. }, 
so it is not so strange that there are many 
pairs of maximizing states $| A(t) \rangle_{\rm{max}}$ and $| B(t) \rangle_{\rm{max}}$ 
allowed by the maximizing principle. 
For the maximizing states the normalized matrix element 
$\langle \hat{\cal O} \rangle_Q^{BA}$ is evaluated 
and time-develops in the same way as Eqs.(\ref{OBAmaxtilde}) and (\ref{ddtOAtildeAtildeQ}).

\subsubsection{Introduction of the $Q$-Hermitian coordinate and momentum operators: $\hat{q}_{Q}$ and $\hat{p}_{Q}$}

To consider concrete examples of $\langle \hat{\cal O} \rangle_Q^{\tilde{A} \tilde{A}}$, 
let us define $Q$-Hermitian coordinate and momentum operators 
$\hat{q}_{Q,a}$ and $\hat{p}_{Q,b}$ by 
\begin{eqnarray}
\hat{q}_{Q,a} &\equiv& \frac{a}{2} \left( \hat{q}_\mathrm{new} +  \hat{q}_\mathrm{new}^{\dag_Q} \right)  
=
a e^{i \frac{\theta}{2}} \cos{ \frac{\theta}{2}} \hat{q}_\mathrm{new} , \label{qhatQa} \\  
\hat{p}_{Q,b} &\equiv& \frac{b}{2} \left( \hat{p}_\mathrm{new} +  \hat{p}_\mathrm{new}^{\dag_Q} \right) 
=
b e^{-i \frac{\theta}{2}} \cos{ \frac{\theta}{2}} \hat{p}_\mathrm{new},  \label{phatQb}
\end{eqnarray}
where $a$ and $b$ are real parameters that are properly chosen. 
In the second equalities of Eqs.(\ref{qhatQa}) and (\ref{phatQb}) we have used 
Eqs.(\ref{Q-1qhatnewdagQ}) and (\ref{Q-1phatnewdagQ}), respectively. 
$\hat{q}_{Q,a}$ and $\hat{p}_{Q,b}$ obey the commutation relation 
$[\hat{q}_{Q,a},  \hat{p}_{Q,b} ] = a b  i \hbar \cos^2{\frac{\theta}{2}}$. 
We are interested in introducing $Q$-Hermitian coordinate and momentum operators 
that obey the same commutation relation as the usual one. 
So let us choose $a=b= \frac{1}{\cos{\frac{\theta}{2}}}$ symmetrically, and define 
$\hat{q}_{Q}$ and $\hat{p}_{Q}$ by 
\begin{eqnarray}
\hat{q}_{Q} 
&\equiv& 
\hat{q}_{Q, \frac{1}{\cos{\frac{\theta}{2}}}} 
= e^{i \frac{\theta}{2}}  \hat{q}_\mathrm{new} , \label{qhatQ}   \\  
\hat{p}_{Q} 
&\equiv&
\hat{p}_{Q,\frac{1}{\cos{\frac{\theta}{2}}}} 
= 
e^{-i \frac{\theta}{2}} \hat{p}_\mathrm{new},  \label{phatQ}
\end{eqnarray}
so that they satisfy the commutation relation $[\hat{q}_{Q} , \hat{p}_{Q}]= i \hbar$.

Naively Eq.(\ref{qhatQ}) looks strange if one wants to consider 
eigenstates for the two supposedly identical operators. 
In fact, $\hat{q}_{Q}$ is Hermitian with regard to the modified inner product $I_Q$, 
and thus has only real eigenvalues, which, though, do not have 
eigenstates belonging to the (true) Hilbert space for 
$I_Q$, the $Q$-Hilbert space $\cal{H}_Q$. 
Rather, $\hat{q}_{Q}$ has only delta-function-normalizable eigenstates 
with regard to $I_Q$, which means that these eigenstates for $\hat{q}_{Q}$ 
belong to an extension of $\cal{H}_Q$ 
by completion in the weak topology for it. 
Now it is a priori -- and indeed it is so -- possible 
that such eigenstates belonging to the extension of $\cal{H}_Q$ could even be true 
Hilbert space vectors under a different inner product such as the usual inner product $I$. 
Therefore, Eq.(\ref{qhatQ}) is not, as it looks at first, contradictory, 
even if we note that $e^{i \frac{\theta}{2}}  \hat{q}_\mathrm{new}$ on the right-hand side 
has all complex numbers $q$ as left-hand eigenvalues in the sense of the 
Hermitian conjugate of Eq.(\ref{qhatqket=qqket_new}) being 
${}_m\langle_\mathrm{new}~ q | \hat{q}_\mathrm{new} ={}_m\langle_\mathrm{new}~ q | q$, 
and that $\hat{q}_\mathrm{new}$ has no right-hand eigenvalues at all 
on the (true) Hilbert space for 
the usual inner product $I$, not even on the extension of it. 
Extension using the inner products $I_Q$ and $I$ does not lead to 
the same space of extended vectors. 
These seeming problems will be discussed further in our subsequent paper\cite{KNHBN_ho2}.

\subsubsection{Hamiltonian expressed in terms of $Q$-Hermitian coordinate and momentum operators}

In order to formulate the $Q$-Hermitian Hamiltonian $\hat{H}_{Qh}$ 
in terms of $Q$-Hermitian 
coordinate and momentum operators $\hat{q}_{Q}$ and $\hat{p}_{Q}$,  
we rewrite the Hamiltonian $\hat{H}$ in Eq.(\ref{hoHamiltonian}) as 
\begin{eqnarray}
\hat{H} 
&=&
\frac{e^{i \theta}}{2m} \hat{p}_Q^2 +\frac{m {\omega}^2 e^{-i \theta}}{2}  \hat{q}_Q^2 
=\frac{\hat{p}_Q^2}{2m'}+\frac{1}{2}m' {\omega}^2 \hat{q}_Q^2, 
\end{eqnarray}
where we have introduced $m' \equiv r_m e^{-i\theta_\omega}$. 
Then, since $\hat{H}^{\dag^Q}$ is given by 
\begin{eqnarray}
\hat{H}^{\dag^Q} 
&=&
\frac{e^{-i \theta}}{2m^* } \hat{p}_Q^2 +\frac{m^* {\omega^*}^2 e^{i \theta}}{2}  \hat{q}_Q^2 
=
\frac{\hat{p}_Q^2}{2{m'}^*}+\frac{1}{2}{m'}^* {{\omega}^*}^2 \hat{q}_Q^2, 
\end{eqnarray}
the $Q$-Hermitian part of $\hat{H}$, $\hat{H}_{Qh}=\frac{1}{2} \left( \hat{H} +  \hat{H}^{\dag^Q}  \right)$, 
is given by 
\begin{eqnarray}
\hat{H}_{Qh}
&=& 
\cos{\theta_\omega}
\left[ 
\frac{1}{2 r_m } \hat{p}_Q^2 
+\frac{r r_\omega }{2}  \hat{q}_Q^2 \right] 
=
\frac{\hat{p}_Q^2}{2m_h} + \frac{1}{2}m_h {\omega_h}^2 \hat{q}_Q^2, \label{HQh}
\end{eqnarray}
where we have introduced 
\begin{eqnarray}
&&m_h \equiv \frac{|m'|^2}{\text{Re}m'} = \frac{r_m}{\cos{\theta_\omega}} , 
\label{m_h} \\
&& 
\omega_h \equiv \frac{\sqrt{\text{Re}m'  \text{Re}(m' {\omega}^2) }}{|m'|} 
= 
r_\omega \cos{\theta_\omega} . \label{omega_h}
\end{eqnarray}
Similarly, the anti-$Q$-Hermitian part of $\hat{H}$, 
$\hat{H}_{Qa}=\frac{1}{2} \left( \hat{H} -  \hat{H}^{\dag^Q}  \right)$, is given by 
\begin{eqnarray}
\hat{H}_{Qa}
&=& i \sin{\theta_\omega} \left[ 
\frac{1}{2 r_m } \hat{p}_Q^2 
+\frac{r r_\omega}{2}  \hat{q}_Q^2 \right]  
=
-i\left[ \frac{\hat{p}_Q^2}{2m_a}+\frac{1}{2}m_a {\omega_a}^2 \hat{q}_Q^2  \right] , \label{HQa}
\end{eqnarray}
where we have introduced 
\begin{eqnarray}
&&m_a \equiv \frac{|m'|^2}{\text{Im}m'} 
= -\frac{r_m }{\sin{\theta_\omega}}, 
\label{m_a} \\
&&\omega_a \equiv \frac{\sqrt{\text{Re}m'  \text{Re}(m' {\omega}^2) }}{|m'|} 
= - r_\omega \sin{\theta_\omega}. \label{omega_a}
\end{eqnarray} 

To check the consistency, let us see  
the other expression for $\hat{H}$ given by Eq.(\ref{H_hbaromega_n}). 
Since $\hat{H}^{\dag^Q}$ is given by Eq.(\ref{HdagQ}), we obtain 
$\hat{H}_{Qh} 
= \hbar r_\omega \cos{\theta_\omega}  \left(\hat{n}_1 + \frac{1}{2} \right)$ and 
$\hat{H}_{Qa} 
= i\hbar r_\omega \sin{\theta_\omega}  \left(\hat{n}_1 + \frac{1}{2} \right)$, 
which lead to 
\begin{equation}
\hat{H}_{Qa}  
= i \tan{\theta_\omega} \hat{H}_{Qh} . \label{HQa_itanthetaomega_HQh}
\end{equation}
Considering Eqs.(\ref{HQh}) and (\ref{HQa}), we obtain 
\begin{eqnarray}
&&m_h = - \tan{\theta_\omega} m_a  , \label{m_h_-tanthetaomega_m_a} \\
&&(m_h \omega_h)^2 = (m_a \omega_a)^2 .  \label{m_h_omega_h2}
\end{eqnarray}
We find that Eqs.(\ref{HQh}) and (\ref{HQa}) satisfy Eq.(\ref{HQa_itanthetaomega_HQh}), 
and that Eqs.(\ref{m_h}), (\ref{m_a}), (\ref{omega_h}), and (\ref{omega_a}) obey 
Eqs.(\ref{m_h_-tanthetaomega_m_a}) and (\ref{m_h_omega_h2}), 
so they are consistent.

\subsubsection{The classical solution to the harmonic oscillator model}

In the generic case where $\text{Im} \omega < 0$, 
we evaluate $\langle \hat{q}_Q \rangle_Q^{\tilde{A} \tilde{A}}$ and 
$\langle \hat{p}_Q \rangle_Q^{\tilde{A} \tilde{A}}$. 
$\langle \hat{q}_Q \rangle_Q^{\tilde{A} \tilde{A}}$ is given by 
\begin{eqnarray} 
\langle \hat{q}_Q \rangle_Q^{\tilde{A} \tilde{A}} 
&=& |a(T_A)|^2 {}_1\langle 0 |_Q \hat{q}_Q | 0 \rangle_1  
\nonumber \\ 
&\propto& {}_1\langle 0 |_Q (\hat{a}_1 + \hat{a}_2^\dag) | 0 \rangle_1  
\nonumber \\ 
&=&0 , \label{qQhatAtildeAtildeQ=0} 
\end{eqnarray}
where in the second line we have used Eqs.(\ref{qhatQ}) and (\ref{newqhat}), 
and in the last equality we have utilized 
Eqs.(\ref{a10ket1=0}), (\ref{a20ket2=0}), and (\ref{Qdef}). 
Similarly, $\langle \hat{p}_Q \rangle_Q^{\tilde{A} \tilde{A}}$ is given by 
\begin{eqnarray} 
\langle \hat{p}_Q \rangle_Q^{\tilde{A} \tilde{A}} 
&=& |a(T_A)|^2 {}_1\langle 0 |_Q \hat{p}_Q | 0 \rangle_1  \nonumber \\ 
&\propto& {}_1\langle 0 |_Q (\hat{a}_1 - \hat{a}_2^\dag) | 0 \rangle_1  \nonumber \\ 
&=&0 , \label{pQhatAtildeAtildeQ=0}
\end{eqnarray}
where in the second line we have used Eqs.(\ref{phatQ}) and (\ref{newphat}), 
and in the last equality we have utilized 
Eqs.(\ref{a10ket1=0}), (\ref{a20ket2=0}), and (\ref{Qdef}). 
In addition, Eq.(\ref{ddtOAtildeAtildeQ}) for $\hat{\cal O}$ being $\hat{q}_Q$ or $\hat{p}_Q$ 
is expressed as 
\begin{eqnarray}
\frac{d}{dt} \langle \hat{q}_Q \rangle_Q^{\tilde{A} \tilde{A}} 
&=&
\frac{1}{m_h} \langle \hat{p}_Q \rangle_Q^{\tilde{A} \tilde{A}}  
=0 , \label{ddtqQhatAtildeAtildeQ}  \\
\frac{d}{dt} \langle \hat{p}_Q \rangle_Q^{\tilde{A} \tilde{A}} 
&=&
- m_h \omega_h^2 \langle \hat{q}_Q \rangle_Q^{\tilde{A} \tilde{A}}  
=0 , \label{ddtpQhatAtildeAtildeQ}
\end{eqnarray}
where in the second equalities of Eqs.(\ref{ddtqQhatAtildeAtildeQ}) 
and (\ref{ddtpQhatAtildeAtildeQ}) we have used Eqs.(\ref{pQhatAtildeAtildeQ=0}) 
and (\ref{qQhatAtildeAtildeQ=0}), respectively. 
Combining Eqs.(\ref{ddtqQhatAtildeAtildeQ}) and (\ref{ddtpQhatAtildeAtildeQ}), 
we obtain the classical equation of motion: 
\begin{eqnarray}
m_h \frac{d^2}{dt^2} \langle \hat{q}_Q \rangle_Q^{\tilde{A} \tilde{A}} 
&=&
- m_h \omega_h^2 \langle \hat{q}_Q \rangle_Q^{\tilde{A} \tilde{A}}  
=0 . \label{d2dt2QQhatAtildeAtildeQ}
\end{eqnarray}
Thus the generic classical solution to the harmonic oscillator model is just zero, 
as shown in the above relations.

In the special case where $\text{Im} \omega = 0$ we do no have 
specific solutions, but only have the relations between 
$\langle \hat{q}_Q \rangle_Q^{A_{\rm{max}} A_{\rm{max}}}$ and 
$\langle \hat{p}_Q \rangle_Q^{A_{\rm{max}} A_{\rm{max}}}$: 
\begin{eqnarray}
\frac{d}{dt} \langle \hat{q}_Q \rangle_Q^{A_{\rm{max}} A_{\rm{max}}} 
&=&
\frac{1}{m_h} \langle \hat{p}_Q \rangle_Q^{A_{\rm{max}} A_{\rm{max}}}, 
\label{ddtqQhatAtildeAtildeQ2}  \\
\frac{d}{dt} \langle \hat{p}_Q \rangle_Q^{A_{\rm{max}} A_{\rm{max}}} 
&=&
- m_h \omega_h^2 \langle \hat{q}_Q \rangle_Q^{\tilde{A} \tilde{A}}, 
\label{ddtpQhatAtildeAtildeQ2}
\end{eqnarray}
which lead to the classical equation of motion:  
\begin{eqnarray}
m_h \frac{d^2}{dt^2} \langle \hat{q}_Q \rangle_Q^{\tilde{A} \tilde{A}} 
&=&
- m_h \omega_h^2 \langle \hat{q}_Q \rangle_Q^{\tilde{A} \tilde{A}}. 
\label{d2dt2QQhatAtildeAtildeQ2}
\end{eqnarray}
Our model in this case is almost the same as the harmonic oscillators in the RAT 
in the sense that there are no imaginary parts of the eigenvalues for the Hamiltonian.  
Hence we cannot specify the classical solution 
unless we are additionally given an initial (or final) condition.

\section{Discussion} \label{sec:discussion}

In the future-included CAT we have formulated and studied the harmonic oscillator model 
defined with a mass $m$ and an angular frequency $\omega$ 
that are taken to be complex numbers. 
Utilizing the complex coordinate formalism~\cite{Nagao:2011za}, 
we defined the Hamiltonian $\hat{H}$ for the harmonic oscillator model. 
For the model to be reasonable we need some 
restrictions on $m$ and $\omega$. 
We found that, according to the argument of $m$ and $\omega$, the model is 
classified into several different theories, and drew the phase diagram.  
Except for at the two corners representing inverted harmonic oscillators in the RAT, 
we formulated two pairs of annihilation and creation operators and 
two series of eigenstates $|n \rangle_1$ and $|n \rangle_2$ 
for the Hamiltonians $\hat{H}$ and $\hat{H}^\dag$ respectively, 
with several algebraically elegant properties as seen in the usual harmonic oscillator in the RAT. 
Our eigenstates $| n \rangle_1$ and $| n \rangle_2$ are not normalized in the usual sense, 
but are $Q$-normalized, i.e. 
normalized in the modified inner product $I_Q$, with respect to which 
the eigenstates of the Hamiltonian $\hat{H}$ become orthogonal to each other. 
In addition, we constructed coherent states.

Furthermore, we applied to the harmonic oscillator model the maximization 
principle~\cite{Nagao:2015bya, Nagao:2017cpl, Nagao:2017book, Nagao:2017ztx}, 
which is the main assumption used by a theorem of ours presented in Sect.~\ref{sec:application_of_maximization_principle}. 
The theorem states that, 
provided that an operator $\hat{\cal O}$ is $Q$-Hermitian, i.e. 
Hermitian with respect to the modified inner product $I_Q$, 
the normalized matrix element (weak value) 
$\langle \hat{\cal O} \rangle_Q^{BA}$ defined in Eq.(\ref{OQBA}) 
becomes real and time-develops under a $Q$-Hermitian Hamiltonian 
for the past and future states selected such that 
the absolute value of the transition amplitude from 
the past state to the future state is maximized. 
In the RAT, coherent states describe classical physics nicely. 
So, as a preliminary study, 
supposing that $|A(T_A) \rangle$ and $|B(T_B) \rangle$ are given by coherent states, 
we evaluated   $\langle \hat{q}_\mathrm{new} \rangle_Q^{BA}$ 
and $\langle \hat{p}_\mathrm{new} \rangle_Q^{BA}$,  
and obtained a nice classical equation of motion. 
This suggests that if we obtain a real observable $\langle \hat{\cal O} \rangle_Q^{B_{\rm{max}} A_{\rm{max}}}$ for the maximizing states via the maximization principle, 
then a nice classical solution is realized. 
Incidentally, introducing $Q$-Hermitian coordinate and momentum operators 
$\hat{q}_Q$ and $\hat{p}_Q$, and rewriting the Hamiltonian  $\hat{H}$ 
in terms of $\hat{q}_Q$ and $\hat{p}_Q$,  
we found that we can obtain via the maximization principle 
an effective theory that is described by the $Q$-Hermitian 
Hamiltonian expressed in terms of $\hat{q}_Q$ and $\hat{p}_Q$. 
However, we have finally obtained 
via the maximization principle the ground state as the generic solution to the harmonic oscillator 
model.  
This might be a somewhat tedious result, but what does this imply?
In our universe, every kind of oscillation can be approximately regarded as 
a harmonic oscillator near the bottom of each potential. 
Therefore, if we suppose that our harmonic oscillator model describes our universe, 
then our solution of the ground state would be very natural. 
In addition, if the universe consists of a lot of approximate harmonic oscillators,
we would see all unexcited except for the few that happened to be almost the RAT. 
We should also point out that we obtained a real-valued solution, 
because $\langle \hat{q}_Q \rangle_Q^{\tilde{A} \tilde{A}} =0 \in \mathbf{R}$ and 
$\langle \hat{p}_Q \rangle_Q^{\tilde{A} \tilde{A}} =0 \in \mathbf{R}$. 
Furthermore, it is interesting that we obtained the $Q$-Hermitian Hamiltonian 
that is expressed in terms of 
$Q$-Hermitian coordinate and momentum operators.

What should we study next? 
In this paper we studied the harmonic oscillator model 
except for at the two corners in the phase diagram in Fig.\ref{fig:phase_diagram_ho}.  
So it is very important to study this model 
in the limit at these corners representing inverted harmonic oscillators in the RAT. 
In particular, inverted harmonic oscillators would be very interesting to study, 
at least from the point of view of regarding such an inverted harmonic oscillator 
as a typically simplified inflaton potential for the slow roll inflation in the early universe. 
Also, it is interesting to investigate the concrete expression for $Q$ in the harmonic 
oscillator model. 
Furthermore, in this paper 
we studied the harmonic oscillator model by utilizing the maximization principle, 
where $|A(T_A) \rangle$ and $|B(T_B) \rangle$ 
are $Q$-normalized, i.e. normalized in the modified inner product $I_Q$. 
On the other hand, it is also important to investigate the model where 
$|A(T_A) \rangle$ and $|B(T_B) \rangle$ are normalized in the usual inner product $I$. 
Such a theory is more complicated to study,  
because we cannot fully utilize the orthogonality of the eigenstates of the Hamiltonian $\hat{H}$. 
Due to this difficulty, we have not yet studied in general 
such a version of the maximization principle. 
However, it would be easier to study it in a concrete model such as the harmonic oscillator.  
We would like to report on such studies in the future.

\section*{Acknowledgments}

Many parts of this work were 
accomplished during the authors' stays in Volosko, Croatia, in the summer of 2012. 
K.N. would like to thank Klara Pavicic for her special kindness in arranging his trip there, 
and the members and visitors of NBI for their kind hospitality 
during his visits to Copenhagen. He was supported in part by 
Grant-in-Aid for Scientific Research (No.21740157) 
from the Ministry of Education, Culture, Sports, Science and Technology (MEXT, Japan). 
H.B.N. is grateful to NBI for allowing him to work at the institute as emeritus. 
In addition, the authors would like to show their gratitude to Ivan and Maja Arcanin 
for their kind hospitality during their stays in Volosko, 
and dedicate this work to the soul of Ivan Arcanin.

\appendix

\section{Detail study of the classification of our harmonic oscillator model by $m$ and $\omega$}\label{appendix:classification}

In this appendix, based on the argument in Sect.~\ref{subsubsec:principle_interpretation}, 
we present an explicit study of the classification of our harmonic oscillator model 
according to the values of $\theta_m$ and $\theta_\omega$. 
This enables us to draw the phase diagram in Fig.\ref{fig:phase_diagram_ho}, which is shown in 
Sect.~\ref{subsubsec:phase_diagram}.

\subsection{ The $0 \le \theta_m < \frac{\pi}{2} $ case }

In this case, since $\cos \theta_m>0$ 
the real part of the mass $m$, $m_\mathrm{R}=r_m \cos \theta_m$, is positive.\footnote{In particular, 
for the $\theta_m =0$ case, $m$ is the real positive mass: $m=r_m$.} 
We choose $a=1$ in Eq.(\ref{mnewam}). 
The quantum Hamiltonian $\hat{H}$ is given by Eqs.(\ref{hoHamiltonian}) and (\ref{ho_potential_hat}), 
and $| A(t) \rangle $ and $| B(t) \rangle $ 
time-develop according to Eqs.(\ref{Atket}) and (\ref{Btket}). 
So let us call this the usual time theory (UTT). 
Based on the signs of $V_\mathrm{R}$ and $V_\mathrm{I}$ 
we can identify the theory as a harmonic oscillator (HO), a free particle, 
or an inverted harmonic oscillator (IHO).

The five regions classified below Eq.(\ref{V_Imomega22}) 
are interpreted as follows: 
\begin{enumerate}
\item For $\theta_\omega =  -\frac{\theta_m}{2}$:

$V_\mathrm{R} > 0$, $V_\mathrm{I}=0$, so this is a harmonic oscillator (HO).

\item For $-\frac{\theta_m}{2} - \frac{\pi}{4} < \theta_\omega <
  -\frac{\theta_m}{2}$:

$V_\mathrm{R} > 0$, $V_\mathrm{I}<0$, so this is a harmonic oscillator (HO).

\item For $\theta_\omega=-\frac{\theta_m}{2} - \frac{\pi}{4}$: 

$V_\mathrm{R} = 0$, $V_\mathrm{I}<0$, so this is a free particle with an imaginary potential. 

\item For $-\frac{\theta_m}{2} - \frac{\pi}{2} <  \theta_\omega < -\frac{\theta_m}{2} - \frac{\pi}{4}$:

$V_\mathrm{R} < 0$, $V_\mathrm{I}<0$, so this is an inverted harmonic oscillator (IHO).

\item For $\theta_\omega=-\frac{\theta_m}{2} - \frac{\pi}{2}$:

$V_\mathrm{R} < 0$, $V_\mathrm{I}=0$, so this is an inverted harmonic oscillator (IHO).
\end{enumerate}

\subsection{ The $\theta_m = \frac{\pi}{2} $ case }

In this case, since $e^{i\theta_m}=i$, the mass $m$ is purely imaginary: 
$m=i r_m$. Since $m_\mathrm{I}=r_m>0$, we choose $a=-i$ in Eq.(\ref{mnewam}), 
and introduce a new mass $\tilde{m}$ by 
$\tilde{m} \equiv -im = r_m$, 
so that the real part of the new mass $\tilde{m}$ becomes positive. 
Let us define purely imaginary times by 
$\tilde{t} \equiv -it$, $\tilde{T}_A \equiv -i T_A$, $\tilde{T}_B \equiv -i T_B$, 
and another angular frequency by 
$\tilde\omega \equiv i \omega$, 
so that $\omega t = \tilde\omega \tilde{t}$. 
Then the coordinate and momentum are rewritten as 
$q(t) = q(i\tilde{t}) \equiv \tilde{q}(\tilde{t})$ and 
$p(t)= m \dot{q}(t) = \tilde{m} \dot{ \tilde{q} } (\tilde{t}) \equiv \tilde{p} (\tilde{t})$, 
where we have introduced $\tilde{q}(\tilde{t})$ and $\tilde{p} (\tilde{t})$ and used the relation 
$\dot{q}(t) = -i \frac{d}{d\tilde{t}} \tilde{q}(\tilde{t}) = -i \dot{ \tilde{q} } (\tilde{t})$. 
Using these new quantities and variables, we can rewrite the classical 
Hamiltonian as 
$H = \frac{p^2}{2m} + \frac{1}{2} m \omega^2 q^2 
= - i \tilde{H}_{\tilde{m}, \tilde{\omega}}$, 
where we have introduced 
$\tilde{H}_{\tilde{m}, \tilde{\omega}} \equiv \frac{\tilde{p}^2}{2 \tilde{m}} 
+\tilde{V}$ and  
$\tilde{V} \equiv \frac{1}{2} \tilde{m} \tilde{\omega}^2 \tilde{q}(\tilde{t})^2$. 
Then its quantum Hamiltonian is given by 
$\hat{\tilde{H}}_{\tilde{m}, \tilde{\omega}} 
\equiv \frac{\hat{p}^2}{2 \tilde{m}} 
+\hat{\tilde{V}}$, where 
$\hat{\tilde{V}} \equiv\frac{1}{2} \tilde{m} \tilde{\omega}^2 \hat{q}^2$. 
$| \tilde{A}(\tilde{t}) \rangle \equiv | A(t) \rangle $ and 
$| \tilde{B}(\tilde{t}) \rangle \equiv | B(t) \rangle $ 
time-develop according to $| \tilde{A}(\tilde{t}) \rangle 
= e^{-\frac{i}{\hbar} \hat{\tilde{H}}_{\tilde{m}, \tilde{\omega}} (\tilde{t}-\tilde{T}_A) }  
| \tilde{A}(\tilde{T}_A) \rangle$ and  
$| \tilde{B}(\tilde{t}) \rangle 
= e^{-\frac{i}{\hbar} \hat{\tilde{H}}_{\tilde{m}, \tilde{\omega} }^\dag (\tilde{t}-\tilde{T}_B) }  
| \tilde{B}(\tilde{T}_B) \rangle$, respectively. 
Thus, in the present case, our theory can be identified as 
the imaginary time theory (ITT) defined with the Hamiltonian 
$\hat{\tilde{H}}_{\tilde{m}, \tilde{\omega}}$.

Using the relations $\text{Re}\tilde{V} = \text{Re}(iV) =- V_\mathrm{I}$ and 
$\text{Im}\tilde{V} = \text{Im}(iV) =V_\mathrm{R}$, 
we interpret the five regions classified below Eq.(\ref{V_Imomega22}) as follows: 
\begin{enumerate} 
\item For $\theta_\omega =  -\frac{\theta_m}{2}  \quad\Leftrightarrow\quad \theta_\omega =  -\frac{\pi}{4}$:

$\text{Re}\tilde{V} = 0$, $\text{Im}\tilde{V}  > 0$, so this is a free particle with an imaginary potential. 

\item For $-\frac{\theta_m}{2} - \frac{\pi}{4} < \theta_\omega <
  -\frac{\theta_m}{2}  \quad\Leftrightarrow\quad - \frac{\pi}{2} < \theta_\omega <
  -\frac{\pi}{4}$:

$\text{Re}\tilde{V}  >0$, $\text{Im}\tilde{V}  > 0$, so this is a harmonic oscillator (HO).

\item For $\theta_\omega=-\frac{\theta_m}{2} - \frac{\pi}{4} \quad\Leftrightarrow\quad \theta_\omega=- \frac{\pi}{2}$: 

$\text{Re}\tilde{V}  > 0$, $\text{Im}\tilde{V} = 0$, so this is a harmonic oscillator (HO). 

\item For $-\frac{\theta_m}{2} - \frac{\pi}{2} <  \theta_\omega < -\frac{\theta_m}{2} - \frac{\pi}{4} \quad\Leftrightarrow\quad - \frac{3}{4}\pi <  \theta_\omega < -\frac{\pi}{2}$:

$\text{Re}\tilde{V}  > 0$, $\text{Im}\tilde{V}  < 0$, so this is a harmonic oscillator (HO).

\item For $\theta_\omega=-\frac{\theta_m}{2} - \frac{\pi}{2} \quad\Leftrightarrow\quad \theta_\omega=- \frac{3}{4}\pi$: 

$\text{Re}\tilde{V} = 0$, $\text{Im}\tilde{V} < 0$, so this is a free particle with an imaginary potential. 
\end{enumerate}

\subsection{ The $\frac{\pi}{2} < \theta_m \leq \pi$ case }

In this case, since $\cos \theta_m<0$, 
the real part of the mass $m$, $m_\mathrm{R}= r_m \cos \theta_m$, 
is negative.\footnote{In particular, 
for the $\theta_m = \pi$ case, $m$ is the real negative mass: $m=-r_m$.} 
In a sensible theory the real part of the mass should be positive. 
So we choose $a=-1$ in Eq.(\ref{mnewam}), 
and introduce a flipped mass $m'$ by $m' \equiv - m$, 
so that the real part of $m'$ is positive. 
Let us define flipped times by 
$t' \equiv -t$, $T'_A \equiv - T_A$ and $T'_B \equiv - T_B$, 
and also a flipped angular frequency $\omega'$ by $\omega' \equiv - \omega$, 
so that $\omega t = \omega' t'$. 
Then the coordinate and momentum are rewritten as 
$q(t)=q(-t') \equiv q'(t')$ and $p(t)=m \dot{q}(t) = m'  \dot{q}'(t') \equiv p'(t')$, 
where we have introduced $q'(t')$ and $p'(t')$, and used the relation 
$\dot{q}(t) = - \frac{d}{dt'} q'(t') =- \dot{q}'(t')$. 
In terms of such flipped quantities and new variables 
the classical Hamiltonian is expressed as $H = - H'_{m', \omega'}$, 
where $H'_{m', \omega}$ is defined by 
$H'_{m', \omega'} (q', p') \equiv \frac{{p'}^2}{2m'} + V'$ and 
$V' \equiv \frac{1}{2} m' {\omega'}^2 {q'(t')}^2$. 
Its quantum Hamiltonian is given by 
$\hat{H'}_{m', \omega'} 
\equiv \frac{\hat{p}^2}{2m'}
+\hat{ V'}$, where 
$\hat{ V'} \equiv \frac{1}{2} m' {\omega'}^2 \hat{q}^2$. 
$|A'(t') \rangle \equiv | A(t) \rangle$ and $|B'(t') \rangle \equiv | B(t) \rangle$ 
time-develop according to 
$| A'(t') \rangle 
= e^{-\frac{i}{\hbar} \hat{H'}_{m', \omega'} (t'-T'_A)} | A'(T'_A) \rangle$ and 
$| B'(t') \rangle 
= e^{-\frac{i}{\hbar}\hat{H'}_{m', \omega'}^\dag (t'-T'_B)} | B'(T'_B) \rangle$, respectively. 
Our theory in the present case can be identified as 
the flipped time theory (FTT), where the state $| A'(T'_A) \rangle$ 
time-develops backward from the future time $T_A'$ to 
the past time $T'_B$, 
while another state $| B'(T'_B) \rangle$ time-develops forward 
from the past time $T_B'$ to the future time $T'_A$.

Using the relations $\text{Re}V' = - V_\mathrm{R}$ and $\text{Im}V' =  - V_\mathrm{I}$, 
we interpret the five regions classified below Eq.(\ref{V_Imomega22}) as follows: 
\begin{enumerate}
\item For $\theta_\omega =  -\frac{\theta_m}{2}$:

$\text{Re}V' < 0$, $\text{Im}V' = 0$, so this is an inverted harmonic oscillator (IHO).

\item For $-\frac{\theta_m}{2} - \frac{\pi}{4} < \theta_\omega <
  -\frac{\theta_m}{2}$:

$\text{Re}V' < 0$, $\text{Im}V' > 0$, so this is an inverted harmonic oscillator (IHO). 

\item For $\theta_\omega=-\frac{\theta_m}{2} - \frac{\pi}{4}$: 

$\text{Re}V' = 0$, $\text{Im}V' > 0$, so this is a free particle with an imaginary potential. 

\item For $-\frac{\theta_m}{2} - \frac{\pi}{2} <  \theta_\omega < -\frac{\theta_m}{2} - \frac{\pi}{4}$:

$\text{Re}V' > 0$, $\text{Im}V' > 0$, so this is a harmonic oscillator (HO).

\item For $\theta_\omega=-\frac{\theta_m}{2} - \frac{\pi}{2}$: 

$\text{Re}V' > 0$, $\text{Im}V' = 0$, so this is a harmonic oscillator (HO).
\end{enumerate}

%
%
\section{Explicit expressions for our ground states $| 0 \rangle_1$ and $| 0 \rangle_2$}\label{appendix:ground_states}

In this appendix, to complement our definition of $| n \rangle_1$ and $| n \rangle_2$ 
in Sect.~\ref{subsec:norm_nket1_nket2}, 
we present explicit expressions for our ground states $| 0 \rangle_1$ and $| 0 \rangle_2$.

To show the definition of our ground states $| 0 \rangle_1$ and $| 0 \rangle_2$ 
explicitly, utilizing the definitions of $\hat{q}_\mathrm{new}$ and 
$\hat{p}_\mathrm{new}$ given in Eqs.(\ref{def_qhat_new}) and (\ref{def_phat_new}), 
we rewrite $\hat{a}_1$ and $\hat{a}_2$ given in Eqs.(\ref{a_1}) and (\ref{a_2}) 
in terms of $\hat{q}$ and $\hat{p}$ as 
\begin{eqnarray}
&&\hat{a}_1 = 
\sqrt{\frac{m\omega}{2 \hbar}} 
\frac{1}{\sqrt{1- \epsilon \epsilon'}}
\left( 1 - \frac{\epsilon'}{m\omega} \right) 
\left(\hat{q} +\frac{i\hat{p} }{(m\omega)_1} \right) ,
\label{a_1_old_expression} \\
&&\hat{a}_2 = 
\sqrt{\frac{m^* \omega^*}{2 \hbar}} 
\frac{1}{\sqrt{1- \epsilon \epsilon'}}
\left( 1 + \frac{\epsilon'}{m^* \omega^*} \right) 
\left(\hat{q} +\frac{i\hat{p} }{(m \omega)_2^*} \right), \label{a_2_old_expression}
\end{eqnarray}
where $(m\omega)_1$ and $(m\omega)_2$ are defined by 
\begin{eqnarray}
&&(m\omega)_1 \equiv 
\frac{m\omega - \epsilon'}{1 - m\omega \epsilon}, \label{m_omega_1} \\
&&(m\omega)_2 \equiv 
\frac{m\omega + \epsilon'}{1 + m\omega \epsilon}. \label{m_omega_2}  
\end{eqnarray}
Then, operating $\langle q |$ on Eqs.(\ref{a10ket1=0}) and (\ref{a20ket2=0}), we obtain 
\begin{eqnarray}
&&\left( q + \frac{\hbar}{(m \omega)_1} \frac{\partial}{\partial q} \right) \langle q | 0 \rangle_1=0, \\
&&\left( q + \frac{\hbar}{(m \omega)_2^*} \frac{\partial}{\partial q} \right) \langle q | 0 \rangle_2=0. 
\end{eqnarray}
Thus the real $q$ representations of our ground states are expressed as 
\begin{eqnarray}
&&\langle q | 0 \rangle_1 = C_1 \exp\left( - \frac{(m \omega)_1}{2\hbar} q^2 \right), 
\label{qbrq0ket1_momega1} \\
&&\langle q | 0 \rangle_2 = C_2^* \exp\left( - \frac{(m \omega)_2^*}{2\hbar} q^2 \right), 
\label{qbrq0ket2_momega2*} 
\end{eqnarray}
where $C_1$ and $C_2$ are normalization factors to be determined by Eq.(\ref{2branmket1}). 
For $\langle q | 0 \rangle_1$ and $\langle q | 0 \rangle_2$ to be convergent, 
we need the conditions 
\begin{eqnarray}
&&\text{Re} ( m\omega )_1 >0, \label{real_momega1>0} \\
&&\text{Re} ( m\omega )_2 >0, \label{real_momega2>0} 
\end{eqnarray}
respectively. 
Hence, for the convergence of both $\langle q | 0 \rangle_1$ and $\langle q | 0 \rangle_2$, 
remembering Eqs.(\ref{m_omega_1}) and (\ref{m_omega_2}), 
we must assume $\epsilon' < \text{Re} \left( m\omega \right) < \frac{1}{\epsilon}$. 
For small $\epsilon$ and $\epsilon'$, this is essentially 
$\text{Re} ( m\omega ) > 0$, which is equivalent to the condition in Eq.(\ref{momegatheta_cond}).

To determine the normalization factors $C_1$ and $C_2$ by Eq.(\ref{2branmket1}), 
let us evaluate ${}_2\langle 0 | 0 \rangle_1$ as follows: 
\begin{eqnarray}
{}_2\langle 0 | 0 \rangle_1 
&=& \int dq ~{}_2\langle 0 | q \rangle \langle q | 0 \rangle_1 \nonumber \\
&=& C_1 C_2 \int dq \exp\left( - \frac{(m \omega)_1 + (m \omega)_2 }{2\hbar} q^2 \right) \nonumber \\
&=& C_1 C_2 \sqrt{ \frac{\pi\hbar (1 - m^2 \omega^2 \epsilon^2)}{ m \omega (1 -\epsilon \epsilon') } },  
\end{eqnarray}
where in the second line the convergent condition for the integral  
\begin{equation}
\text{Re}\left\{ ( m\omega )_1 + ( m\omega )_2 \right\} >0 \label{real_momega1+momega2>0} 
\end{equation}
is automatically satisfied under the conditions 
in Eqs.(\ref{real_momega1>0}) and (\ref{real_momega2>0})\footnote{The convergence of 
${}_2\langle 0 | 0 \rangle_1$, i.e. 
$|{}_2\langle 0 | 0 \rangle_1 | < \infty$, is also obtained under the convergence of both 
${}_1\langle 0 | 0 \rangle_1$ and ${}_2\langle 0 | 0 \rangle_2$ by utilizing the Schwarz inequality: 
$|{}_2\langle 0 | 0 \rangle_1 | \le \sqrt{{}_1\langle 0 | 0 \rangle_1 ~{}_2\langle 0 | 0 \rangle_2 }$. }, 
which become Eq.(\ref{momegatheta_cond}) for small $\epsilon$ and $\epsilon'$. 
We choose symmetrically 
\begin{equation}
C_1=C_2= \left\{ \frac{ m \omega (1 -\epsilon \epsilon') }{\pi\hbar (1 - m^2 \omega^2 \epsilon^2)} \right\}^{\frac{1}{4}} \equiv C , \label{C_1=C_2=C}
\end{equation}
so that ${}_2\langle 0 | 0 \rangle_1=1$. 
Thus our ground states $| 0 \rangle_1$ and $| 0 \rangle_2$ are specified by 
Eqs.(\ref{qbrq0ket1_momega1}), (\ref{qbrq0ket2_momega2*}), and (\ref{C_1=C_2=C}).

Incidentally, we give the explicit expressions for our excited states $| n \rangle_1$ and $| n \rangle_2$ 
for our reference. 
Substituting Eqs.(\ref{a_2_old_expression}) and (\ref{a_1_old_expression}) 
for Eqs.(\ref{nket1normalized}) and (\ref{nket2normalized}), respectively, and operating $\langle q|$ on them, 
we obtain the real $q$ representations of $| n \rangle_1$ and $| n \rangle_2$ as follows:  
\begin{eqnarray}
&&\langle q | n \rangle_1 
= \frac{C}{\sqrt{n !}} 
\left\{
\sqrt{\frac{m\omega}{2 \hbar}} 
\frac{1}{\sqrt{1- \epsilon \epsilon'}} 
\left( 1 + \frac{\epsilon'}{m\omega} \right)
\right\}^n 
\left( q - \frac{\hbar}{(m\omega)_2} \frac{\partial}{\partial q} \right)^n 
\exp\left( - \frac{(m \omega)_1}{2\hbar} q^2 \right), \label{nket1normalized_oldqp} \\
&&\langle q | n \rangle_2 
= \frac{C^*}{\sqrt{n !}} 
\left\{
\sqrt{\frac{m^* \omega^*}{2 \hbar}} 
\frac{1}{\sqrt{1- \epsilon \epsilon'}} 
\left( 1 - \frac{\epsilon'}{m^* \omega^*} \right)
\right\}^n 
\left( q - \frac{\hbar}{(m\omega)_1^*} \frac{\partial}{\partial q} \right)^n 
\exp\left( - \frac{(m \omega)_2^*}{2\hbar} q^2 \right), \label{nket2normalized_oldqp} \nonumber \\ 
\end{eqnarray}
where $C$, $(m\omega)_1$, and $(m\omega)_2$ are given in 
Eqs.(\ref{C_1=C_2=C}), (\ref{m_omega_1}), and (\ref{m_omega_2}).


\end{document}